\documentclass[twocolumn, journal]{IEEEtran}

\usepackage{booktabs} 
\usepackage{multirow}
\usepackage{mathtools}

\usepackage{amssymb}

\usepackage{enumitem}

\usepackage{smartdiagram}
\usesmartdiagramlibrary{additions}

\graphicspath{{Figs/}}

\usepackage{pdfpages}
\usepackage{pdflscape}
\usepackage{amsmath}
\usepackage{amsthm}

\DeclareMathOperator*{\argmin}{arg\,min}

\usepackage{amsfonts}

\usepackage{color}

\usepackage{setspace}
\usepackage{enumitem}
\usepackage{xcolor}

\usepackage{bm}
\usepackage{bbm}

\DeclareMathOperator{\Tr}{Tr}

\usepackage{tikz}
\usepackage{pgf}
\usetikzlibrary{arrows,automata}
\usetikzlibrary{shapes}
\tikzstyle{data44}=[rectangle split,rectangle split parts=2,draw,text centered]

\usepackage{subfigure}

\DeclareMathAlphabet{\mathcal}{OMS}{cmsy}{m}{n}

\usetikzlibrary{matrix}

\tikzset{
  BarreStyle/.style =   {opacity=.3,line width=14 mm,color=#1},
  node style ge/.style={},
  node style sp/.style={},
  yl/.style={},
  arrow style mul/.style={},
}

\newtheorem{proposition}{Proposition}

\newtheorem{remark}{Remark}
 
\newtheorem{definition}{Definition}

\newtheorem{probstate}{Problem Statement}

\usepackage{threeparttable,booktabs}

\usepackage[colorlinks,
            linkcolor=blue,
            anchorcolor=blue,
            citecolor=blue
            ]{hyperref}

\usepackage{cite}

\usepackage{mdwlist}

\usepackage{algorithm}
\usepackage{algorithmicx}
\usepackage{algpseudocode}

\errorcontextlines\maxdimen

\makeatletter
    \newcommand*{\algrule}[1][\algorithmicindent]{\makebox[#1][l]{\hspace*{.5em}\thealgruleextra\vrule height \thealgruleheight depth \thealgruledepth}}%
\newcommand*{\thealgruleextra}{}
\newcommand*{\thealgruleheight}{.75\baselineskip}
\newcommand*{\thealgruledepth}{.25\baselineskip}

\newcount\ALG@printindent@tempcnta
\def\ALG@printindent{%
    \ifnum \theALG@nested>0
        \ifx\ALG@text\ALG@x@notext
        \else
            \unskip
            \addvspace{-1pt}
            \ALG@printindent@tempcnta=1
            \loop
                \algrule[\csname ALG@ind@\the\ALG@printindent@tempcnta\endcsname]%
                \advance \ALG@printindent@tempcnta 1
            \ifnum \ALG@printindent@tempcnta<\numexpr\theALG@nested+1\relax
            \repeat
        \fi
    \fi
    }%
\usepackage{etoolbox}
\patchcmd{\ALG@doentity}{\noindent\hskip\ALG@tlm}{\ALG@printindent}{}{\errmessage{failed to patch}}
\makeatother

\newbox\statebox
\newcommand{\myState}[1]{%
    \setbox\statebox=\vbox{#1}%
    \edef\thealgruleheight{\dimexpr \the\ht\statebox+1pt\relax}%
    \edef\thealgruledepth{\dimexpr \the\dp\statebox+1pt\relax}%
    \ifdim\thealgruleheight<.75\baselineskip
        \def\thealgruleheight{\dimexpr .75\baselineskip+1pt\relax}%
    \fi
    \ifdim\thealgruledepth<.25\baselineskip
        \def\thealgruledepth{\dimexpr .25\baselineskip+1pt\relax}%
    \fi
    \State #1%
    \def\thealgruleheight{\dimexpr .75\baselineskip+1pt\relax}%
    \def\thealgruledepth{\dimexpr .25\baselineskip+1pt\relax}%
}

\newcommand{\hig}[1]{\textcolor{black}{#1}}

\usepackage{textcomp}

\usepackage{stmaryrd}

\usepackage{comment}

\usepackage{graphicx,subfigure}

\usepackage{stfloats}
\usepackage{tabularx }

\newcommand\smallmath[2]{#1{\raisebox{\dimexpr \fontdimen 22 \textfont 2
      - \fontdimen 22 \scriptscriptfont 2 \relax}{$\scriptscriptstyle #2$}}} 
\newcommand\smallotimes{\smallmath\mathbin\otimes}

\usepackage{makecell}

\usepackage{colortbl}

\newcolumntype{P}[1]{>{\centering\arraybackslash}p{0.66cm}}
 
\usepackage{anyfontsize}
\usepackage{t1enc}

\newcommand{\HP}{^{\circ 2}}
\newcommand{\HR}{^{\circ \frac{1}{2}}}
\newcommand{\T}{^{\scriptscriptstyle\rm T}}
\newcommand{\B}{^{\scriptscriptstyle\rm L}}  
\newcommand{\U}{^{\scriptscriptstyle\rm U}}
\newcommand{\D}{^{\diamond}}

\usepackage[strict]{changepage}
\usepackage{relsize}

\begin{document}

\title{Bumpless Topology Transition}

\author{Tong Han,\IEEEmembership{}
        Yue Song,~\IEEEmembership{Member, IEEE,} and 
        David J. Hill,~\IEEEmembership{Life Fellow, IEEE}
\thanks{This work was supported by the Research Grants Council of the Hong Kong Special Administrative Region through the General Research Fund under Project No. 17209419.}
\thanks{T. Han and Y. Song are with the Department of Electrical and Electronic Engineering, University of Hong Kong, Hong Kong (e-mail: hantong@eee.hku.hk; yuesong@eee.hku.hk).}
\thanks{D. J. Hill is with the Department of Electrical and Electronic Engineering, The University of Hong Kong, Hong Kong, and also with the School of Electrical Engineering and Telecommunications, The University of New SouthWales, Kensington, NSW 2052, Australia (e-mail: dhill@eee.hku.hk).}

}

\maketitle

\begin{abstract}
    The topology transition problem of transmission networks is becoming increasingly crucial with topological flexibility more widely leveraged to promote high renewable penetration. 
    This paper proposes a novel methodology to address this problem. Aiming at achieving a bumpless topology transition regarding both static and dynamic performance, this methodology utilizes various eligible control resources in transmission networks to cooperate with the optimization of line-switching sequence. Mathematically, a composite formulation is developed to efficiently yield bumpless transition schemes with AC feasibility and stability both ensured. With linearization of all non-convexities involved and tractable bumpiness metrics, a convex mixed-integer program firstly optimizes the line-switching sequence and partial control resources. Then, two nonlinear programs recover AC feasibility, and optimize the remaining control resources by minimizing the $\mathcal{H}_2$-norm of associated linearized systems, respectively. The final transition scheme is selected by accurate evaluation including stability verification using time-domain simulations. 
    Finally, numerical studies demonstrate the effectiveness and superiority of the proposed methodology to achieve bumpless topology transition. 
\end{abstract}
 

\IEEEpeerreviewmaketitle

\section*{Notation, Acronym and Nomenclature}

{
\normalsize{$\!\!\!\!\!\!\!\!$
\textit{Notation}:$\!$ 
    \textit{(1)} For a vector $\bm{x}$, $\bm{x}\D$ is the diagonal matrix with entries of $\bm{x}$ on the main diagonal. For a square matrix $\bm{X}$,  $\bm{X}\D$ is the vector of the main diagonal of $\bm{X}$. 
    \textit{(2)} For a vector $\bm{x}$, $x_i$ is the entry of $\bm{x}$ associated with object $i$. For example, for bus voltages $\bm{v}$, $v_i$ is the voltage of bus $i$.
    \textit{(3)} Domains of most constants and variables are ignored. Unless otherwise specified, bold lowercase letters are vectors in $\mathbb{R}$ with proper dimension, normal lowercase letters are scalars in $\mathbb{R}$.
    \textit{(4)} $\Vert \!\cdot\!\Vert_{2, \bm{w}}$ denotes the 2-norm of a vector with $\bm{w}$ being the weight vector. 
    \textit{(5)} $\overline{1, n}$ denotes the set of integers from 1 to $n$.
    \textit{(6)} $B\backslash A$ is the set difference of $\!B$ and $\!A$, and $\!|A|\!$ the cardinality of set $\!A$.
}
}

\setlist[description]{labelindent=0pt,style=multiline,leftmargin=2cm}
\begin{description}
    \item[$\mathrm{DVC, SVC}$] Dynamic/static VAR compensator 
    \item[$\mathrm{STATCOM}$]  Static synchronous compensator
    \item[$\mathrm{TCSC}$]  Thyristor-controlled series compensation
    \item[$\mathrm{SG, CIG}$]  Synchronous/converter-interfaced generator 
    \item[$\mathrm{ESS, IM}$] Energy storage system, induction motor
    \item[$\mathrm{SS, VSC}$] Steady state, voltage source converter
\end{description}

\setlist[description]{labelindent=0pt,style=multiline,leftmargin=1.5cm}
\begin{description}
\item[$n_{\rm n}, n_{\rm e}$] Number of buses/branches.
\item[$n_{\rm t}$] A value determining the number of breakpoints in linearization of the network power flow model.

\item[$x_{{\rm q}, i}$] The $q$-axis synchronous resistance of SG $i$. 
\item[$x_i'', x_{{\rm m}, i}''$] Subtransient resistances of SG $i$ and the IM component of load $i$. 
\item[$r_{{\rm s}, i}$] Stator resistance of the IM component of load $i$.
\item[$r_{{\rm c}, i}$] RLC filter resistance of CIG $i$.
\item[$x_{\!\!\!~{\rm cl}\!\!\!~,\!\!\!~ i}, \!x_{\!\!\!~{\rm cc}\!\!\!~,\!\!\!~ i}$]  Fundamental reactances of the RLC filter resistance and capacitance of CIG $i$.   
\item[$v_{0, i}$] Reference voltage of load $i$.
\item[$\alpha_{{\rm p},i}, \alpha_{{\rm q},i}$] Weights$\!$ of constant impedance$\!$ components $\!$of load $i$.
\item[$\beta_{{\rm p},i}, \beta_{{\rm q},i}$] Weights of constant current components of load $i$. 
\item[$\gamma_{{\rm p},i}, \gamma_{{\rm q},i}$] Weights of constant power components of load $i$. 
\item[$\epsilon_{{\rm p},i}, \epsilon_{{\rm q},i}$] Weights of IM components of load $i$. 
\item[$p_{~\!\!\!{\rm{d_{0}}}\!, i},\! q_{~\!\!\!{\rm{d_0}}\!, i}$] Active and reactive power of load $i$ at the steady state where $v_j = v_{0,i}$ with $j = \mathcal{C}(i)$.
\item[$x_{{\rm svg}, i}$] Equivalent reactance between STATCOM $i$ and bus $\mathcal{C}(i)$.
\item[$b_{{\rm b}, e}\B, b_{{\rm b}, e}\U$] Lower/upper bound of the susceptance of branch $e$.
\item[$b_{{\rm b}, e}^0$]  A value of the susceptance of branch $e$ that is able to be taken for any value of $n_{\rm t}$.
\item[$\mathcal{C}(i)$] The bus connected with element $i$.

\item[$\bm{g}_{\rm b} \!+\! j\bm{b}_{\rm b}$] Vector of branch admittances.
\item[$\bm{g}_{\rm lc} \!+\! j \bm{b}_{\rm lc}$] Vector of half ground admittance of branches contributed by line charges.
\item[$\bm{v} \angle \bm{\theta}$] Voltages of buses.
\item[$\bm{b}_{\rm b, tcsc}$] Subvector of $\bm{b}_{\rm b}$ associated with lines with TCSC.
\item[$\bm{p}_{\rm fb}, \bm{q}_{\rm fb}$] Active/reactive powers at the starting buses of branches.
\item[$\bm{p}_{\rm tb}, \bm{q}_{\rm tb}$] Active/reactive powers at the end buses of branches.
\item[$\bm{e} \angle \bm{\delta}$] Electromotive force (emf) of SGs.
\item[$\bm{e}_{\rm s} \angle \bm{\delta}_{\rm s}$] Subtransient emf of SGs.
\item[$\bm{v}_{\rm m} \angle \bm{\theta}_{\rm m}$] Modulation voltages at the outputs of CIGs.
\item[$\bm{p}_{\rm g, ess}$] Subvector of $\bm{p}_{\rm g}$ associated with CIGs with ESS.
\item[$\bm{v}_{\rm g}$]  Subvector of $\bm{v}$ associated with generator buses.
\item[$\bm{p}_{\rm g}, \bm{q}_{\rm g}$] Active/reactive power outputs of generators.
\item[$\bm{e}_{\rm m}\angle \bm{\delta}_{\rm m}$] Internal voltages behind the subtransient impedances of the IM component of loads.
\item[$\bm{v}_{~\!\!\!\rm s~\!\!\!v~\!\!\!g} ~\!\!\angle~\!\! \bm{\theta}_{\!\rm s~\!\!\!v~\!\!\!g}$] Modulation voltages at the outputs of STATCOMs.
\item[$\bm{v}_{\rm dvc}$] Subvector of $\bm{v}$ associated with buses with DVCs.
\item[$\bm{\epsilon}_{\rm p}, \bm{\epsilon}_{\rm q}$] Vector of all $\epsilon_{{\rm p}, i}$, vector of all $\epsilon_{{\rm q}, i}$
\item[$\bm{q}_{\rm c}$] Reactive power outputs of DVCs. 
\item[$\bm{b}_{\rm svc}$] Susceptances of SVCs. 
\item[$\bm{p}_{\rm go}$]  Subvector of $\bm{p}_{\rm g}$ associated with generators excluding converter-based generator with ESSs.
\item[$\bm{p}_{\rm gs}$] Subvector of $\bm{p}_{\rm g}$ associated with SGs.
\item[$\bm{p}_{\rm gc}, \bm{q}_{\rm gc}$] Subvector of $\bm{p}_{\rm g}$/$\bm{q}_{\rm g}$ associated with CIGs.
\item[$\bm{m}_{\rm cg}, \bm{d}_{\rm cg}$] Inertia and damping coefficients of CIGs. 
\item[$\bm{g}_{\rm s}\D \!+\! j \bm{b}_{\rm s}\D$] Admittance matrix of the network contributed by bus shunts and ground admittances of transformers. 
\item[$\bm{E}, \tilde{\bm{E}}$] Oriented incidence matrix of graph $\mathcal{G}$ with each branch assigned an arbitrary and fixed orientation, and its entry-wise absolute value. 
\item[$\bm{E}_{\rm f}, -\bm{E}_{\rm t}$]  Formed by replacing all -1/1 entries in $\bm{E}$ by 0.


\end{description}


\section{Introduction}

\IEEEPARstart{T}{opological} flexibility of transmission systems should be more fully leveraged to accommodate high penetration of renewable energy \cite{4-970}. Control actions that improve system performance by optimizing transmission network topology are commonly known as transmission switching or optimal transmission switching (OTS) \cite{4-62}. In conventional transmission networks, OTS shows its capability to reduce generation cost \cite{4-62}, improve system stability \cite{4-361}, and etc. OTS is also evolving to consider the features of renewable generation and tackle the challenges posed by high renewable penetration \cite{4-1369, 4-859}. 

Despite different mechanisms, all OTS faces the same topology transition problem, namely how to realize the transition from the initial topology to the target one given by the OTS model \cite{4-1309}. As concluded in \cite{4-1309}, due to more frequent execution of OTS and new dynamic properties associated with the dominance of converters, the topology transition problem becomes increasingly crucial with the transformation to high renewable penetrated transmission networks. Moreover, the necessity of particular topology transition strategies is also proved numerically by the observed violations of operational constraints caused by the ad hoc topology transition \cite{4-1309}.

Nonetheless, studies on the topology transition problem of transmission networks are limited. For switching of a single line, Martins et al. \cite{4-1285} designed a generation rescheduling method to reduce the induced generator rotor shaft impacts. The stability issue associated with transmission switching events was investigated in \cite{4-558, 4-161}. Huang et al. \cite{4-558} revealed the small-disturbance instability triggered by line switching and thus the necessity of some controls to prevent it. Owusu-Mireku and Chiang \cite{4-161} showed that the existence of a steady state power flow solution fails to ensure transient stability after line switching, calling for considerations of system dynamics in the topology transition problem. In contrast to the previous works which addressed the topology transition problem with a single switching action \cite{4-1285, 4-161} or fixed line-switching sequence \cite{4-558}, the authors in \cite{4-1309} proposed the concept of optimal topology transition and developed a mixed-integer program to determine the optimal trajectory of topology transition. As a preliminary solution to the topology transition problem, only the static performance during transition was addressed in \cite{4-1309}.

For microgrids, similar topology transition problems exist to facilitate network reconfiguration. El-Sayed et al. \cite{4-1286} developed a nonconvex mixed-integer nonlinear program (MINLP) which reduces the negative impacts of topology transition on transient voltage by simultaneous optimization of the line-switching sequence and droops of distributed generators. The power flow through the switched lines was used as an indicator of the peak of voltage transients, yielding a tractable objective function. Following the paradigm in \cite{4-1286}, a sensitivity-based method was developed in \cite{4-1287} to improve the computational efficiency. More system details such as three-phase unbalance were considered while the line-switching sequence was fixed.

Toward a more complete and practical methodology to tackle the topology transition problem of transmission networks, the following three aspects should be further addressed:

\textit{(1)} The static and dynamic factors associated with topology transition should be both considered. Firstly, a comprehensive evaluation of transition processes require the metrics capturing both static and transient performance, rather than those with only one of them considered \cite{4-1286, 4-1309}.
Secondly, as a basic requirement, the transition methodology should contain mechanisms to ensure stability of the entire transition process, which however, is neglected in existing works. 

\textit{(2)} Since line switching itself is a large disturbance, only optimizing the line-switching sequence as in \cite{4-1309} is potentially insufficient with system dynamics involved. Thus, we need to seek eligible control resources to stabilize the system such as the generation in \cite{4-1285}. Given that practical topology transition generally contains switching of multiple indefinite lines, these control resources should at least have fast response time and be able to together pose system-wide impacts. 

\textit{(3)} Computation tractability for real transmission networks is indispensable. 
The topology transition problem where line-switching sequence is optimized essentially features a nonconvex MINLP. Unlike for small-sized microgrids in \cite{4-1286}, such a program is computationally intractable for real transmission networks and thus requires particular study. Moreover, the transient components in performance metrics potentially further complicate the solution method. 

Considering the foregoing, this paper develops a novel and powerful methodology for the topology transition problem of transmission networks with the following main contributions:

\textit{(1)} $\!$The topology transition problem is treated as the problem of bumpless topology transition (BTT), where eligible common control resources in transmission networks are utilized to achieve bumpless transition along with optimizing the line-switching sequence. 
These control resources, including terminal voltages of generators, outputs of ESSs, inertia and damping of CIGs, DVCs, TCSCs and line switching, cover all aspects of transmission networks.

\textit{(2)} A metric is developed to comprehensively quantify the magnitude of the response, which we will call  bumpiness, of the transition processes. 
It consists of boundedness and volatility of the steady-state components of performance outputs which capture the global bumpiness, and an integral term of the transient-state components to represent the local bumpiness after each line switching. In addition, two surrogates of the integral term, based on the $\mathcal{H}_2$ norm of associated linearized systems and the instantaneous system state after line switching, are developed for computational tractability. 

\textit{(3)} A composite mathematical formulation of the BTT problem is proposed to efficiently produce bumpless transition schemes where AC feasibility and stability are ensured. It contains four models: a mixed-integer second-order programming (MISOCP) model to optimize the line-switching sequence and the control resources which can impact SSs; two nonlinear programming (NLP) models to recover AC feasibility and optimize inertia and damping of CIGs; and a simulation-based model to select the best scheme based on accurate evaluations.

\section{Problem Description and Process of BTT}

\subsection{Problem Description}\label{sec-pd-a}

The transmission network is firstly represented as an undirected graph $\mathcal{G}(\mathcal{V}, \mathcal{E})$ with $\mathcal{V}$ and $\mathcal{E}$ being the sets of buses and branches, respectively. 
The network topology is parameterized as $\bm{z} \!\!\in\!\! \mathbb{B}^{n_{\rm e}}$ whose entry values of 1 or 0 represent
the associated branches are switched on or off. Let $\bm{a}$ be the vector of certain adjustable electrical properties and dynamic parameters of the system, called auxiliary control variables (ACVs). ACVs are allowed to be adjusted to assist the topology transition.

\begin{figure}[h]
	\centering
	\includegraphics[width=1\linewidth]{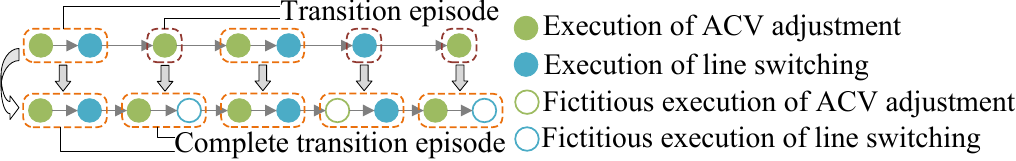} 
	\caption{Illustration of TEs and complete TEs.
    }  
	\label{fig-6-2-6} 
\end{figure}

It is assumed that ACV adjustment and line switching are executed asynchronously, and then the topology transition process is determined by a sequence of executions of ACV adjustment and line switching. 
To represent any transition process as a unique standard form, 
we further introduce the concepts of \textit{complete transition episode} and \textit{transition episode} (TE) given by Definition \ref{def-6-2-0}. They are also illustrated in Fig. \ref{fig-6-2-6} with a transition process. For ease of description and modelling, we introduce some fictitious executions of ACV adjustment and line switching to convert all TEs to complete TEs, as illustrated in Fig. \ref{fig-6-2-6}. \hig{Taking the second TE with only ACV adjustment for example, a fictitious execution of line switching is added for obtaining a complete TE.}

\begin{definition}[Complete TE, TE]\label{def-6-2-0}
    A complete TE is an execution of ACV adjustment and an execution of line switching which follows the former. A TE is a complete TE, or an execution of ACV adjustment or line switching in the transition process excluding all complete TEs.
\end{definition}

We further represent the transition process by the transition trajectory of topology and ACVs. 
Denote the state of topology and ACVs by $(\bm{z}, \bm{a})$. 
Then the $i$-th TE can be denoted as $(\bm{z}^{i\!-\!1}\!, \bm{a}^{i\!-\!1}) \!\!\to\!\! (\bm{z}^{i\!-\!1}\!, \bm{a}^{i}) \!\!\to\!\! (\bm{z}^{i}\!, \bm{a}^{i})$, 
where $(\bm{z}^{i\!-\!1}\!, \bm{a}^{i\!-\!1}) \!\!\to\!\! (\bm{z}^{i\!-\!1}\!, \bm{z}^{i})$ means adjusting $\bm{a}$ from $\bm{a}^{i-1}$ to $\bm{a}^{i}$, and $(\bm{z}^{i-1}\!, \bm{a}^{i}) \!\!\to\!\! (\bm{z}^{i}\!, \bm{a}^{i})$ means changing $\bm{z}$ from $\bm{z}^{i-1}$ to $\bm{z}^i$. 
When $\bm{a}^{i\!-\!1} \!\!=\!\! \bm{a}^i$ or $\bm{z}^{i\!-\!1} \!\!=\!\! \bm{z}^i$, the $i$-th TE contains a fictitious execution of ACV adjustment or line switching. Accordingly, the transition process can be represented by the transition trajectory of topology and ACVs, i.e., $ \cdots \!\to\! (\bm{z}^{i-1}, \bm{a}^{i-1}) \!\!\to\!\! (\bm{z}^{i-1}, \bm{a}^{i}) \!\!\to\!\! (\bm{z}^{i}, \bm{a}^{i}) \!\!\to\!\! \cdots$.

Then the problem of bumpless topology transition focused on in this work is defined by Problem Statement \ref{def-6-2-1}, \hig{where $T$ represents the number of TEs in the transition process.}

\begin{probstate}[Bumpless topology transition]\label{def-6-2-1}
    Given the initial value of $\bm{a} \!\!=\!\! \bm{a}^0$, an initial topology $\bm{z}^0$, and a final topology $\bm{z}^T$, under which the systems are operationally feasible, find a feasible transition trajectory of topology and ACVs, i.e., $(\bm{z}^0, \bm{a}^0) \!\to\! (\bm{z}^0, \bm{a}^1) \!\to\! (\bm{z}^1, \bm{a}^1) \!\to\! (\bm{z}^1, \bm{a}^2) \!\to\! ... \!\to\! (\bm{z}^{T-1}, \bm{a}^{T-1}) \!\to\! (\bm{z}^{T-1}, \bm{a}^{T}) \!\to\! (\bm{z}^{T}, \bm{a}^{T})$ with $\bm{a}^T \!\!=\!\! \bm{a}^0$, such that the transition process is as bumpless as possible.
\end{probstate}

For the selection of ACVs, the following four criteria should be considered: 
\textit{(\romannumeral1) Fast response time.} Since the topology transition process should be as short as possible, ACVs are required to have fast response time. 
\textit{(\romannumeral2) Bumpless.} The purpose of adjusting ACVs is to aid topology transition to be bumpless and therefore the process of adjusting ACVs should be as bumpless as possible, which in general requires ACVs to be continuous. Although benefits of adjusting discrete ACVs for reducing bumps of line switching can outweigh the large bumps caused by the adjustment itself, adjusting both continuous and discrete ACVs complicates the process of topology transition.
\textit{(\romannumeral3) Negligible or low cost.} Negligible or low cost of adjusting ACVs can maximize the profits from topology optimization. 
\textit{(\romannumeral4) Negligible impacts on reliability.} For high VRE penetrated transmission networks, topology transition is potentially frequent, hence frequent adjustment of ACVs. This is required to have negligible impacts on reliability of devices and system operation. 
\textit{(\romannumeral5) System-wide impacts.} Given that line switching actions all over the network can be involved for numerous scenarios of topology transition, the selected ACVs should be able to together pose system-wide impacts.

Table \ref{tab-6-2-1} lists the common control resources in transmission networks. According to their performances regarding the above criteria, at the generation side, terminal voltages of generators, outputs of ESSs, and virtual inertia and damping of CIGs are selected as ACVs. Unlike SGs whose inertia and damping are inherent physical properties, virtual inertia and damping of CIGs are parameters of control loops and thus are tunable \cite{4-1294}. 
At the load side, DVCs, mainly including SVCs and STATCOMs, are selected. In the transmission network, DVCs are commonly used to maintain constant bus voltage and therefore we choose the voltage setpoints of DVCs as ACVs. 
At the network side, TCSC and line switching are selected. Only the open-loop impedance control of TCSC is consider in this work while other control modes can be considered if necessary. Under this control mode, TCSC is equivalent to a constant series reactance operating at its setpoint \cite{4-1303}. Thus, we choose the reactance setpoints of TCSCs as ACVs. 

Line switching used as auxiliary control for topology transition refers to the interim line switching. Let $\mathcal{E}_i \!\subseteq\! \mathcal{E}$ be the set containing all closed lines indicated by $\bm{z}^i$. Specifically, to transition from topology $\!\mathcal{E}_0$ to $\!\mathcal{E}_T$, $|\mathcal{E}_0 \backslash \mathcal{E}_T|$ openings of every line in $\mathcal{E}_0 \backslash \mathcal{E}_T$ and $|\mathcal{E}_T \backslash \mathcal{E}_0|$ closures of every line in $\mathcal{E}_T \backslash \mathcal{E}_0$ are necessary. The others are just interim line switchings, which includes switching of lines not in $(\mathcal{E}_T \backslash \mathcal{E}_0) \cup (\mathcal{E}_0 \!\backslash \mathcal{E}_T)$, switching of lines in $\mathcal{E}_0 \backslash \mathcal{E}_T$ which is not the first open, and switching of lines in $\mathcal{E}_T \!\backslash \mathcal{E}_0$ which is not the first closure. In fact, by criterion \textit{(\romannumeral2)}, line switching is not a candidate for auxiliary control since it is discrete adjustment. However, interim line switching clearly does not complicate the process of topology transition. Accordingly, we incorporated interim line switching into $\bm{z}$ instead of $\bm{a}$, such that all ACVs are continuous variables.

\begin{table*}[ht]
    \caption{Common control resources in transmission networks}\label{tab-6-2-1}
    \centering
    \setlength{\tabcolsep}{0pt} 

    \setlength{\aboverulesep}{-0.8pt}
    \setlength{\belowrulesep}{0.5pt}
    \setlength{\extrarowheight}{-.3ex}
    \small{
    \begin{tabularx}{1.0\textwidth}{p{0.8cm}p{4cm}<{\centering}p{2.6cm}<{\centering}p{1.2cm}<{\centering}p{2cm}<{\centering}p{2.2cm}<{\centering}p{1.8cm}<{\centering}|p{3.2cm}<{\centering}}
    \toprule   
    & \multicolumn{1}{c}{Control resources} & \multicolumn{1}{c}{Response time} & \multicolumn{1}{c}{Bumpless} & \multicolumn{1}{c}{Cost} & \multicolumn{1}{c}{Reliability Impact} & {Competent} & \multirow{10}{3cm}{
        \scriptsize{
            $\!\!\!^{\textcolor{blue}{\rm a}}$ The value is the ramp rate, expressed in the percent of maximum capacity per minute $\!\!\!\!$ or second, that a generator or ESS changes its output. 
            \\
            $^{\textcolor{blue}{\rm b}}$ The two ranges respectively $\!\!\!\!$ correspond to STATCOM and$\!\!\!$ SVC. 
            \\ 
            $^{\textcolor{blue}{\rm c}}$ This depends on the type of OLTC, namely that the tap adjustment is continuous or discrete. 
          }    
    }  
    \\  
    \cmidrule{1-7}
    \parbox[t]{1mm}{\multirow{4}{*}{\rotatebox[origin=c]{90}{\makecell{\footnotesize{Generation}\\\footnotesize{side}}}}} 
    & Outputs of SGs  &  2-30\%/min$^{\textcolor{blue}{\rm a}}$ \cite{4-1293} & Yes & High & Negative & No &  \\
    & Terminal voltage of generators     & $<$1s & Yes & Negligible &   Negligible & Yes  &  \\
    & Outputs of ESSs & $>$200\%/s$^{\textcolor{blue}{\rm a}}$      & Yes & Low  &  Negligible & Yes &    \\
    & Inertia and damping of CIGs & $<$1s & Yes & Negligible & Negligible & Yes &  \\ \cmidrule{1-7}
    \parbox[t]{1mm}{\multirow{4}{*}{\rotatebox[origin=c]{90}{\makecell{~~~\footnotesize{Load}\\~~~\footnotesize{side}}}}} 
    & Load demands              & $<$30s \cite{4-1292} & Yes  & High  & Negligible & No &  \\
    & Shunt capacitors          & $<$1s                & No  & Negligible & Negative & No &   \\
    & Dynamic VAR compensators  & $<$5ms, 20-40ms $^{\textcolor{blue}{\rm b}}$  & Yes & Negligible & Negligible & Yes &   \\ \cmidrule{1-7}
    \parbox[t]{1mm}{\multirow{4}{*}{\rotatebox[origin=c]{90}{\makecell{~~~$\!$\footnotesize{Network$\!$}\\~~~\footnotesize{side}}}}} 
    & On-load tap changer (OLTC)            &  3s-10s \cite{4-1291} & Yes/No$^{\textcolor{blue}{\rm c}}$  & Negligible  & Negative & No &  \\
    & TCSC            & 15-20ms \cite{4-1303} & Yes & Negligible & Negligible   & Yes  &   \\
    & Line switching  & $<$1s  & No &  Negligible & Negligible  & Yes &   \\
    \bottomrule
    \end{tabularx}
    }
\end{table*}

Furthermore, we make the following assumptions: 
\begin{itemize} 
    \item[A.1] In each TE, ACV adjustment is fast enough and the induced dynamic response is smooth enough, such that the associated transient process of system state is negligible for evaluation of bumpiness of the transition process.
    \item[A.2] Line switching is performed asynchronously and thus at most one line is switched in each TE, and line switching and ACV adjustment are both performed after the system reach a SS.
    \item[A.3] Except for ACVs and network topology, other control variables and parameters of the system remain constant during the entire transition process.
    \item[A.4] Let $n_{\rm ad}$ and $n_{\rm us}$ be the number of executions of ACV adjustment and interim line switching actions performed, respectively. They are bounded by
    \begin{equation}\label{eq-6-2-a4}
        n_{\rm ad} T_{\rm ad}  +  n_{\rm us}  T_{\rm ls} \leq T_{\max}
    \end{equation} 
    where $T_{\rm ad}$ and $T_{\rm ls}$ are the estimated increases of transition time caused by an execution of ACV adjustment and an interim line switching action, respectively; and $T_{\max}$ is the maximal allowable increase of transition time by ACV adjustment and interim line switching.
\end{itemize}

\begin{remark}
    A.1 is reasonable given the first two selection criteria for ACVs. In addition, limiting the adjustment range of ACVs in each TE also rationalizes A.1, which will be considered in the later BTT model. 

    A.2 is made considering the feasibility of analysing and executing a topology transition. Although the static performance of topology transition can benefit from simultaneous switching of multiple lines, its dynamic process is much more involved due to uncertainties in communication time and relay operating time \cite{4-1309}. Moreover, switching lines and adjusting ACVs during transients may produce more bumpless topology transition, which however, pose extremely high requirements on real-time performance of monitoring and communication. 

    A.3 can be moderate provided that the duration of the BTT process is relatively short. This duration mainly depends on the number of line switching actions and how rapidly the oscillations decay after a line switching. For medium-scale networks, the number of line switching actions for OTS is generally around 10 \cite{4-63}. For large-scale networks, more line switching actions are performed while the system state changes more steadily. Moreover, rapid oscillation decay after a line switching can be ensured by the objective function of the later BTT model where these oscillations are suppressed. 

    A.4 is reasonable with the purpose to prevent excessive executions of ACV adjustment and interim line switching actions which extend the duration of the BTT process and increase operational complexity immoderately.
\end{remark}

\subsection{Process of BTT}

\begin{figure}[h]
	\centering
	\includegraphics[width=\linewidth]{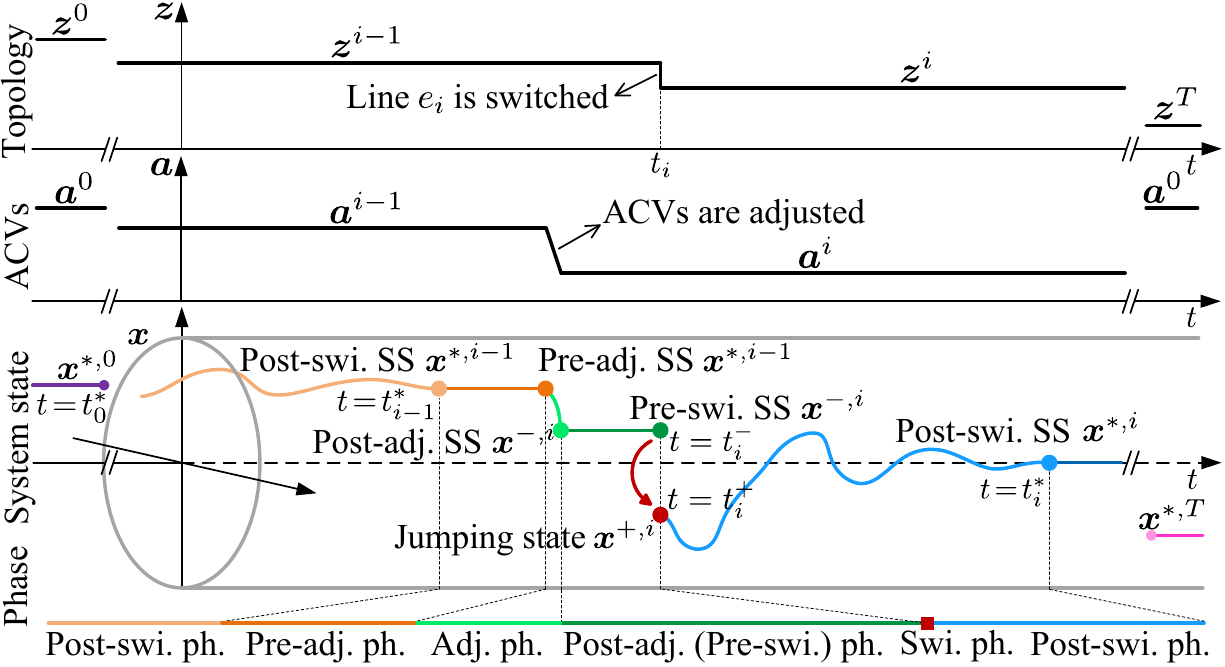} 
	\caption{Illustration of the process of BTT with the $i$-th TE as an example. Here ``adj.'', ``swi.'' and ``ph.'' are abbreviations for ``adjustment'', ``switching'' and ``phase'', respectively.}  
	\label{fig-6-2-1} 
\end{figure}

Fig. \ref{fig-6-2-1} illustrates the process of BTT, \hig{where the top plot is the change of network topology, the middle one the change of ACVs, and the bottom one the evolution of system states together with the phase partition.} 
The process begins at $t \!\!=\!\! t_0^*$, with $(\bm{z}, \bm{a})\!\!=\!\!(\bm{z}^0, \bm{a}^0)$ and $\bm{x} \!=\! \bm{x}^{*,0}$. Taking the $i$-th TE which is a complete one for example, the associated process contains 5 phases as follows: 

\textit{1) The pre-adjustment phase.} This phase is the period after the system reaches a SS $\bm{x}^{*,i-1}$, called post-switching SS, at $t=t_{i-1}^*$, and before ACVs are adjusted. 

\textit{2) The adjustment phase.} This phase is the period where ACVs are adjusted. In this phase, \hig{the values of ACVs are together smoothly adjusted from $\bm{a}^{i-1}$ to $\bm{a}^i$, as illustrated in the middle plot of Fig. \ref{fig-6-2-1}.} Meanwhile, the system state changes from the pre-adjustment SS, which is the same as the post-switching SS by assumption A.3, to a new SS $\bm{x}^{-, i}$, called post-adjustment SS. By assumption A.1, the transient process of this phase is neglected.  

\textit{3) Post-adjustment or pre-switching phase.} This phase is the period after the system reaches the post-adjustment SS and before a certain line, denoted as $e_i$, is switched.

\textit{4) Switching phase.} This phase is the instant $t_i$ where line $e_i$ is switched. In this phase, the system state jumps from the pre-switching SS at $t=t_i^-$, which is the same as the post-adjustment SS by assumption A.3, to another state $\bm{x}^{+, i}$ at $t = t_i^+$, called jumping state. 

\textit{5) Post-switching phase.} This phase is the period after the system jumps to the jumping state and if the system is stable, ending with reaching the next post-switching SS $\bm{x}^{*, i}$ at $t = t_{i}^*$.

The process of BTT ends with the network topology switched to $\bm{z}^T$ and ACVs adjusted to the initial value $\bm{a}^0$. Hereinafter, the system associated with each phase is named with the name of that phase.

\section{System Models}

This section develops the system models used for formulating the BTT problem mathematically. 
First, we partition $\bm{a}$ that contains all the selected ACVs into two subvectors, i.e., $\bm{a}_{\rm s} \!\!=\!\! [\bm{v}_{\rm g}\T, \bm{p}_{\rm g, ess}\T, \bm{v}_{\rm dvc}\T, \bm{b}_{\rm b, tcsc}\T]\T$ consisting of ACVs which can impact SSs, and $\bm{a}_{\rm t} \!\!=\!\! [\bm{m}_{\rm cg}\T, \bm{d}_{\rm cg}\T ]\T$ consisting of ACVs which have an impact on transients but no impact on SSs. Let $\bm{a}_{\rm s}^i$ and $\bm{a}_{\rm t}^i$ be the values of $\bm{a}_{\rm s}$ and $\bm{a}_{\rm t}$ corresponding to $\bm{a}^i$, respectively.

\subsection{Power Flow Models}


\subsubsection{Network} The AC power flow model of the network written in vector form is given by
\begin{adjustwidth}{0em}{}
\begin{subequations}\label{eq-6-2-1}
    \begin{align}
        &\!\!\!\!\!\!
        \begin{bmatrix}
            \! \bm{p}_{\rm fb} \!\!  \\
            \! \bm{p}_{\rm tb} \!\!  \\
            \! \bm{q}_{\rm fb} \!\!  \\
            \! \bm{q}_{\rm tb} \!\!
        \end{bmatrix}
        \!\!\!=\!\!\!
        \left(\!\!
        \begin{bmatrix}
            \!\bm{g}_{\rm b}\D      \!\bm{E}_{\rm f}\T \\
            \!\bm{g}_{\rm b}\D      \!\bm{E}_{\rm t}\T\!\! \\
            \!\!-\!\bm{b}_{\rm b}\D \!\bm{E}_{\rm f}\T\!\! \\
            \!\!-\!\bm{b}_{\rm b}\D \!\bm{E}_{\rm t}\T\!\!
        \end{bmatrix}
        \!\!\! \bm{v}^{2}
        \!\!\!-\!\!\!
        \begin{bmatrix}
            \! \bm{g}_{\rm b}\D      \!\!&\!\! \bm{b}_{\rm b}\D     \!\!\!\\
            \! \bm{g}_{\rm b}\D      \!\!&\!\!\!\!- \!\bm{b}_{\rm b}\D \!\! \\
            \! -\!\bm{b}_{\rm b}\D \!\!&\!\! \bm{g}_{\rm b}\D     \!\!\!\\
            \! -\!\bm{b}_{\rm b}\D \!\!&\!\!\!- \bm{g}_{\rm b}\D \!\!
        \end{bmatrix}\!\!\!\!
        \begin{bmatrix}
            \!\cos(\! \bm{E}\T \! \bm{\theta} \!) \!\! \\
            \!\sin(\! \bm{E}\T \! \bm{\theta} \!) \!\!
        \end{bmatrix}
        \!\!\!\circ\!\!
        \bm{(}\!\bm{1}_4 \!\!\smallotimes\!\! 
        \psi(\!\bm{v}\!) \!\bm{)}
        \!\!\! \right) 
        \!\!\!\circ\!\! (\! \bm{1}_{4} \!\!\smallotimes\!\! \bm{z} )     \label{eq-3-6-ac-power-1:1} 
        \\
        &\!\!\!\!\!\! \begin{bmatrix}
            \!\bm{E}_{\rm g} \bm{p}_{\rm g}  \!\!-\!\! \bm{E}_{\rm d} \bm{p}_{\rm d} \!\! \\
            \!\bm{E}_{\rm g} \bm{q}_{\rm g}  \!\!-\!\! \bm{E}_{\rm d} \bm{q}_{\rm d} \!\!+\!\! \bm{E}_{\rm c} \bm{q}_{\rm c} \!\! 
        \end{bmatrix}
        \!\!\!-\!\!\!
        \begin{bmatrix}
            \! \bm{E}_{\rm f} \bm{p}_{\rm fb}  \!\!+\!\! \bm{E}_{\rm t} \bm{p}_{\rm tb} \!\! \\
            \! \bm{E}_{\rm f} \bm{q}_{\rm fb}  \!\!+\!\! \bm{E}_{\rm t} \bm{q}_{\rm tb} \!\!
        \end{bmatrix}
        \!\!\!=\! \!\! 
        \begin{bmatrix}
            \! \tilde{\bm{E}} \bm{g}_{\rm lc}\D  \bm{z}   \!+\! \bm{g}_{\rm s}  \!\!\\
            \! - \tilde{\bm{E}} \bm{b}_{\rm lc}\D \bm{z} \!-\! \bm{b}_{\rm s}  \!\!
        \end{bmatrix}
        \!\!\!\circ\!\!\!
        \begin{bmatrix}
            \! \bm{v}^{2}\! \\
            \! \bm{v}^{2}\! 
        \end{bmatrix}\!\!\!\!\!\!\!\!
         \label{eq-3-6-ac-power-1:2}    
    \end{align} 
\end{subequations}
\end{adjustwidth}
where $\psi( \bm{v} ) \!=\!\! (\bm{E}_{\rm f}\T \bm{v}) \circ (\bm{E}_{\rm t}\T \bm{v})$, \hig{(\ref{eq-3-6-ac-power-1:1}) models branch power flow, and (\ref{eq-3-6-ac-power-1:2}) models power balance of nodes. Model (\ref{eq-6-2-1}) is obtained by reformulating the common AC power flow equations with voltages in polar coordinates to separate the power flow terms related to different admittance components, and considering the branch status parameterized by $\bm{z}$.}

\subsubsection{Generation}

For CIGs, we assume that they are all VSCs. Then the power injection of CIG $i$ expressed with voltage $v_{{\rm m},i} \angle \theta_{{\rm m},i}$ is give by
\begin{equation}\label{eq-6-2-3}
    \begin{bmatrix}
        p_{{\rm g}, i} \\
        q_{{\rm g}, i}
    \end{bmatrix}
    \!=\!
    v_{{\rm m}, i} v_j 
    \begin{bmatrix}
        g_{{\rm c}, i} &   b_{{\rm c}, i} \\
        b_{{\rm c}, i} &   g_{{\rm c}, i}
    \end{bmatrix}\!
    \begin{bmatrix}
        \cos(\theta_{{\rm m}, i}\!-\!\theta_j) \\
        \sin(\theta_{{\rm m}, i}\!-\!\theta_j)
    \end{bmatrix}
    \!-\!
    \begin{bmatrix}
        \! g_{{\rm c}, i}  \!\\
        \! b_{{\rm cc}, i} \!
    \end{bmatrix} 
    \! v_j^2 
\end{equation}
with $j \!\!=\!\! \mathcal{C}(i)$, $g_{{\rm c}, i} \!\!=\!\! r_{{\rm c}, i} \cdot ({r_{{\rm c}, i}^2 \!\!+\!\! x_{{\rm cl}, i}^2 })^{\!-1}$, 
$b_{{\rm c}, i} \!\!=\!\! x_{{\rm cl}, i} \cdot ({r_{{\rm c}, i}^2 \!+\! x_{{\rm cl}, i}^2 })^{-1}$, 
and $b_{\!{\rm cc}, i} \!\!=\!\! b_{\!{\rm c}, i} + x_{\!{\rm cc}, i}^{-1} $. For simplification purposes, the right-hand side of (\ref{eq-6-2-3}) is denoted as
$\mathcal{F}(v_{{\rm m}, i}, v_j, \theta_{{\rm m}, i}, \theta_j | g_{{\rm c}, i}, b_{{\rm c}, i}, b_{{\rm cc}, i})$.

For SG $i$, its power injection is associated with emf $e_i \angle \delta_i $ and subtransient emf ${e}_{{\rm s}, i} \angle {\delta}_{{\rm s}, i}$ as
\begin{equation}\label{eq-6-2-2-1}
    [p_{{\rm g}, i}~ q_{{\rm g}, i}]\T = \mathcal{F}(e_i, v_j, \delta_i, \theta_j |0, x_{{\rm q}, i}^{-1}, x_{{\rm q}, i}^{-1})
\end{equation}
\begin{equation}\label{eq-6-2-2}
    [p_{{\rm g}, i}~ q_{{\rm g}, i}]\T = \mathcal{F}(e_{{\rm s}, i}, v_j, \delta_{{\rm s}, i}, \theta_j | 0, 1/x_i'', 1/x_i'')
\end{equation}
with $j \!=\! \mathcal{C}(i)$. Note that (\ref{eq-6-2-2}) uses the approximation that $d$-axis and $q$-axis subtransient reactances of a SG are equal.

\subsubsection{Loads}
Considering that load dynamics are involved in the switching phase, we adopt the ZIP-IM load model, which is commonly used for dynamic studies \cite{4-1289, 4-1108}. Thus the power demand of load $i$ is associated with bus voltage by 
\begin{adjustwidth}{-0.8em}{}
\begin{subequations}\label{eq-6-2-4}
    \begin{align}
        & 
        \begin{bmatrix}
            \!p_{{\rm d}, i}\!\! \\
            \!q_{{\rm d}, i}\!\!
        \end{bmatrix} 
        \!\!\!=\!\!\!
        \begin{bmatrix}
            p_{{\rm{d_0}}, i} [ \alpha_{{\rm p},i} (\frac{v_j}{v_{0,i}})^2 \!\!+\!\!  \beta_{{\rm p},i} \frac{v_j}{v_{0,i}} \!\!+\!\! \gamma_{{\rm p},i} \!\!+\! \epsilon_{{\rm p}, i}] \\
            q_{{\rm{d_0}}, i} [ \alpha_{{\rm q},i} (\frac{v_j}{v_{0,i}})^2 \!\!+\!\!  \beta_{{\rm q},i} \frac{v_j}{v_{0,i}} \!\!+\!\! \gamma_{{\rm q},i} \!\!+\! \epsilon_{{\rm q}, i}] 
        \!\!
        \end{bmatrix}
        \label{eq-6-2-4:1} \\
        &
        ~[\epsilon_{{\rm p},i} p_{{\rm{d_0}}, i}~ \epsilon_{{\rm q},i} q_{{\rm{d_0}}, i}]
        \!\!=\!\! - \mathcal{F}(e_{{\rm m}, i}, v_j, \delta_{{\rm m}, i}, \theta_j | g_{{\rm m}, i}, b_{{\rm m}, i}, b_{{\rm m}, i}) \!\!\! \label{eq-6-2-4:2}
    \end{align}
\end{subequations}
\end{adjustwidth}
with $g_{{\rm m}, i} \!\!=\!\! r_{{\rm s}, i} \!\cdot\! (r_{{\rm s}, i}^2 \!\!+\!\! {x''_{{\rm m}, i}}^{\!\!\!\!2})^{\!-1}$, 
$b_{{\rm m}, i} \!\!=\!\! x''_{{\rm m}, i} \!\cdot\! (r_{{\rm s}, i}^2 \!\!+\!\! {x''_{{\rm m}, i}}^{\!\!\!\!2} )^{\!-1}$, and $j \!\!=\!\! \mathcal{C}(i)$. Model (\ref{eq-6-2-4:1}) is based on the approximation that the IM components operate at the same power, independent of the bus voltage, for all SSs. \hig{It is noted that $\alpha_{{\rm p}, i}$, $\alpha_{{\rm q}, i}$, $\beta_{{\rm p}, i}$, $\beta_{{\rm q}, i}$, $\gamma_{{\rm p}, i}$, $\gamma_{{\rm q}, i}$, $\epsilon_{{\rm p}, i}$, and $\epsilon_{{\rm q}, i}$ are the parameters of the ZIP-IM load model whose values for practical power systems are generally obtained by load parameter identification.} 
For the switching phase, load power of the IM components is given by (\ref{eq-6-2-4:2}) with $\epsilon_{{\rm p}, i}$ and $\epsilon_{{\rm q}, i}$ substituted by their variable counterparts.

For DVCs at the load side, if DVC $i$ is an SVC, its reactive power injection is associated with bus voltage by
\begin{equation}\label{eq-6-2-5}
    q_{{\rm c}, i} =  b_{{\rm svc}, i} v_j^2 \text{~with~} j = \mathcal{C}(i)
\end{equation}
Note that for different SSs, $b_{{\rm svc}, i}$ is a variable depending on the voltage setpoint of SVC. For DVC $i$ being a STATCOM, its power injection is associated with bus voltage by
\begin{equation}\label{eq-6-2-6}
    [0~ q_{{\rm c}, i}]\T = \mathcal{F}(v_{{\rm svg}, i}, v_j, \theta_{{\rm svg}, i}, \theta_j |0, x_{{\rm svg}, i}^{-1}, x_{{\rm svg}, i}^{-1})
\end{equation}

\subsubsection{Compact form} To lighten notations, power flow models (\ref{eq-6-2-1})-(\ref{eq-6-2-6}) are denoted in a descriptor form as
\begin{equation}\label{eq-6-2-cfnonlinear}
    f_{\rm p}( \bm{x}_{\rm p} | \bm{z}, \bm{a}_{\rm s} | \bm{y}_{\rm p}) = \bm{0}  
\end{equation}
where $\bm{x}_{\rm p}$ is the vector of all voltage variables, and $\bm{y}_{\rm p}$ is the vector of observed variables which are needed in the formulation of the BTT problem. Specifically, 
\begin{adjustwidth}{-0.5em}{}
\begin{subequations}
    \begin{align}
        & \bm{x}_{\rm p} \!\!=\!\! [\bm{v}\T\!, \bm{\theta}\T\!, \bm{e}\T\!, \bm{\delta}\T\!, \bm{e}_{\rm s}\T\!, \bm{\delta}_{\rm s}\T\!, \bm{v}_{\rm m}\T, \bm{\theta}_{\rm m}\T, \bm{e}_{\rm m}\T, \bm{\delta}_{\rm m}\T, \bm{v}_{\rm svg}\T, \bm{\theta}_{\rm svg}\T]\T \\ 
        & \bm{y}_{\rm p} \!\!=\!\! [ \bm{p}_{\rm fb}\T, \bm{q}_{\rm fb}\T, \bm{p}_{\rm tb}\T, \bm{q}_{\rm tb}\T, \bm{p}_{\rm g}\T, \bm{q}_{\rm g}\T, \bm{\epsilon}_{\rm p}\T, \bm{\epsilon}_{\rm q}\T, \bm{q}_{\rm c}\T, \bm{b}_{\rm svc}, \bm{b}_{\rm b, tcsc}\T ]\T
    \end{align}
\end{subequations} 
\end{adjustwidth}
 
\subsubsection{Linear form}
For computational tractability, we further linearize the above power flow models, which gives, also in a descriptor form
\begin{equation}\label{eq-6-2-cflinear}
    \tilde{f}_{\rm p}( \bm{x}_{\rm p} | \bm{z}, \bm{a}_{\rm s} | \bm{y}_{\rm p}) \leq \bm{0}
\end{equation}
where $\tilde{f}(\cdot)_{\rm p}$ is linear in terms of $\bm{x}_{\rm p}$, $\bm{z}$, $\bm{a}_{\rm s}$ and $\bm{y}_{\rm p}$. See the appendix for the detailed linearization.

\subsection{Dynamic Models}

The system dynamics are formulated by a state-space descriptor form as
\begin{equation}\label{eq-6-2-13}
    \begin{bmatrix}
        \dot{\bm{x}} \\
        \bm{0}
    \end{bmatrix}
    = 
    \begin{bmatrix}
        f(\bm{x}, \bm{\xi}, \bm{z}, \bm{a}) \\
        g(\bm{x}, \bm{\xi}, \bm{z}, \bm{a})
    \end{bmatrix}
    , 
    \bm{y} = h(\bm{x}, \bm{\xi}, \bm{z}, \bm{a})
\end{equation}
where $\bm{x} \in \mathbb{R}^{n_{\rm x}}$, $\bm{\xi} \in \mathbb{R}^{n_{\rm xi}}$ are the vectors of $n_{\rm x}$ state variables and $n_{\rm xi}$ algebraic variables, respectively; 
and $\bm{y} \in \mathbb{R}^{n_{\rm y}}$ is the vector of $n_{\rm y}$ performance outputs.

The linearized model of (\ref{eq-6-2-13}) around a given SS $(\bm{x},\! \bm{\xi}) \!\!=\!\! (\!\bm{x}^*\!, \bm{\xi}^*\!)$, with $\bm{\xi}$ eliminated and an input term added, is
\begin{equation}\label{eq-6-2-14} 
       \!\!\!\!\!
       \begin{aligned}
        & \Delta \dot{\bm{x}}  \!=\!\! \bm{A}(\!\bm{z},\! \bm{a},\! \bm{x}^*\!,\! \bm{\xi}^*\!) \Delta \bm{x} \!+\!\! \bm{B}(\!\bm{x}^{\Delta}\!) \bm{u} \\ 
        & \Delta \bm{y} \!\!=\!\! \bm{C}(\!\bm{z},\! \bm{a},\! \bm{x}^*,\! \bm{\xi}^*\!)  \Delta \bm{x}   \!\!
       \end{aligned}
\end{equation}
where $\Delta\bm{x} \!\!=\!\! \bm{x} \!-\! \bm{x}^*$, $\Delta\bm{y} \!\!=\! \bm{y} \!-\! h(\bm{x}^*, \bm{\xi}^*, \bm{z}, \bm{a})$, $\bm{B}(\bm{x}^{\Delta}) \in \mathbb{R}^{n_{\rm x} \times n_{\rm x}}$ is the input matrix determined by $\bm{x}^{\Delta} \in \mathbb{R}^{n_{\rm x}}$, $\bm{u} \in \mathbb{R}^{n_{\rm x}}$ is the input vector; and
\begin{equation}
        \!\! \bm{A} \!\!=\!\! \frac{\partial f}{\partial \bm{x}} \!-\! \frac{\partial f}{\partial \bm{\xi}} \!\! \left(\! \frac{\partial g}{\partial \bm{\xi}} \!\right)^{\!\!\!-\!1} \!\!\!\frac{\partial g}{\partial \bm{x}}, 
        \bm{C} \!\!=\!\!  
        \begin{bmatrix}
            \!\!\frac{\partial h}{\partial \bm{x}}\!\!-\!\! \frac{\partial h}{\partial \bm{\xi}}  \!\!\left(\! \frac{\partial g}{\partial \bm{\xi}} \!\right)^{\!\!\!-\!1} \!\!\!\frac{\partial g}{\partial \bm{x}} &\!\!\!\!
            -\! \frac{\partial h}{\partial \bm{\xi}} \!\!\left(\! \frac{\partial g}{\partial \bm{\xi}} \!\right)^{\!\!\!-\!1} \!\!\!\frac{\partial g}{\partial \bm{u}} \!\!
        \end{bmatrix}\!\!\!
\end{equation}
all at $(\bm{x}, \bm{\xi}) \!\!=\!\! (\bm{x}^*\!, \bm{\xi}^*)$. Hereinafter, we use $\Psi(\bm{z}, \bm{a})$ to refer to the system given by (\ref{eq-6-2-13}), and the transfer function of (\ref{eq-6-2-14}), denote as $G(s, \bm{z}, \bm{a}, \bm{x}^*, \bm{\xi}^*, \bm{x}^{\Delta})$, to refer to the system (\ref{eq-6-2-14}). Moreover, denote by $\bm{y}^{*, i}$ and $\bm{\xi}^{*, i}$ the points of $\bm{y}$ and $\bm{\xi}$ corresponding to $\bm{x}^{*, i}$, respectively; and $\bm{y}^{-, i}$, $\bm{y}^{+, i}$, and $\bm{\xi}^{+, i}$ are analogous.

\section{Bumpiness Metric and Its Tractable Surrogates}

This section develops metrics to evaluate how bumpless the transition process is, and their surrogates which are able to be incorporated with mathematical models of the BTT problem.

\subsection{Bumpiness Metric}

\begin{figure*}[ht]
	\centering
	\includegraphics[width=\textwidth]{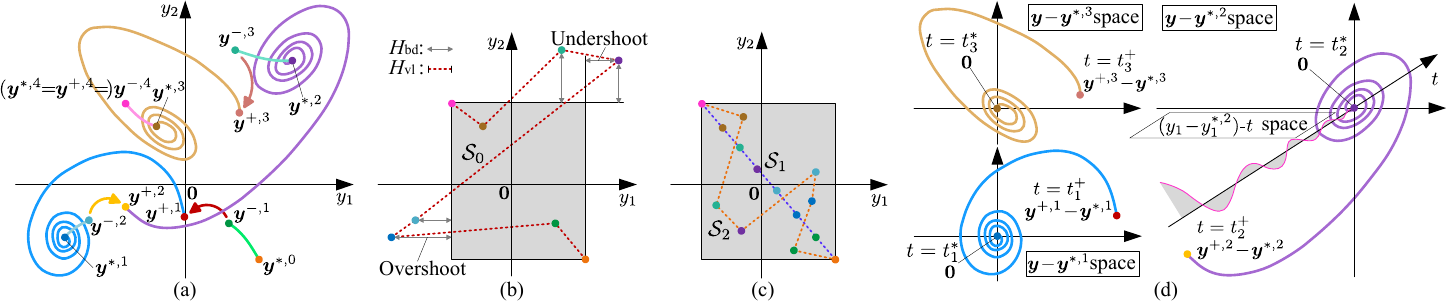} 
	\caption{(a) The schematic trajectory of $\bm{y}$ for a topology transition with $T= 4$. (b)-(d) Illustration of the bumpiness metric for the trajectory of $\bm{y}$ in (a).}
	\label{fig-6-2-23} 
\end{figure*}

We introduce the bumpiness metric using an example of topology transition with $T\!\!=\!4$. The schematic of the trajectory of $\bm{y}$ during the transition is shown in Fig. \ref{fig-6-2-23}-(a). 
The trajectory of $\bm{y}$ is decomposed into two components, i.e., the steady-state component (SSC) as shown in Fig. \ref{fig-6-2-23}-(b), and the transient-state component (TSC) as shown in Fig. \ref{fig-6-2-23}-(d). Bumpiness of the trajectory is analysed following this decomposition. 

\subsubsection{SSC}

Bumpiness of the SSC contains two aspects, called boundedness and volatility. Define the \textit{optimal region} as the intersection of 
all surfaces that pass through $\bm{y}^{*, 0}$ and $\bm{y}^{*,T}$, and are parallel to one axis. \hig{Here $\bm{y}^{*, 0}$ (or $\bm{y}^{*,T}$) can be obtained by solving (\ref{eq-6-2-13}) with $\dot{\bm{x}}=\bm{0}$, $\bm{z}=\bm{z}^{0}$ (or $\bm{z}=\bm{z}^{T}$), $\bm{a}=\bm{a}^{0}$ (or $\bm{a}=\bm{a}^{T}$), or using the solution of power flow model (\ref{eq-6-2-cfnonlinear}) with similar substitution.} 
The boundedness refers to how well the SSC of $\bm{y}$ is bounded by the optimal region. This property is associated with the concepts of overshoot and undershoot. In control theory overshoot is the occurrence of an output exceeding its target and undershoot is the same phenomenon in the opposite direction. Here the target for undershoot is relaxed such that it occurs only when an output falls behind the starting point. 
Overshoot and undershoot (with a slight abuse of terminology, we still use these two term to refer to the similar phenomenons here) of the SSC of $\bm{y}$ of $\mathcal{S}_0$ is illustrated in Fig. \ref{fig-6-2-23}-(b), and the gray rectangle is the optimal region to bound the SSC of $\bm{y}$. It can be seen that better boundedness indicates smaller overshoot and undershoot. Comparatively, the SSCs of $\bm{y}$ of $\mathcal{S}_1$ and $\mathcal{S}_2$ in Fig. \ref{fig-6-2-23}-(c) is bounded by the gray rectangle, with neither overshoot nor undershoot, and thus being more bumpless.

The volatility refers to the degree of number of changes of the SSC of $\bm{y}$ during the topology transition. As shown in Fig. \ref{fig-6-2-23}-(c), the most bumpless topology transition is $\mathcal{S}_1$ where the path of the steady-state points of $\bm{y}$ is the shortest one between $\bm{y}^{*, 0}$ and $\bm{y}^{*, T}$, and all changes of $\bm{y}$ are necessary to realize the topology transition. For the topology transition where the length of this path is longer, such as $\mathcal{S}_0$ and $\mathcal{S}_1$, more and unnecessary changes of the SSC of $\bm{y}$ are involved and thus reducing bumpiness of the topology transition. Thus, deviation between the length of the path of steady-state points of $\bm{y}$ and the shortest one can be an indicator of the volatility.

\subsubsection{TSC}

Boundedness and volatility of the SSC capture the global bumpiness, while the TSC contains information about the local bumpiness after each line switching. Taking the projection of the TSC of $\bm{y}$ from $t=t_2^+$ to $t=t_2^*$, i.e., the pink trajectory in Fig. \ref{fig-6-2-23}-(d) for example, a smaller gray area indicates that the line switching at $t = t_2$ is more bumpless. Therefore, bumpiness of the TSC can be characterized by some analogs of the gray area but in the space of $(\bm{y}-\bm{y}^{*, i})$-$t$. 

According to the above analysis, the proposed bumpiness metric, denoted by $H$, is formally defined as
\begin{equation}\label{eq-6-2-16}
    H((\bm{z}^i, \bm{a}^i | i \!\in\! \overline{1, T})) = H_{\rm bd} + H_{\rm vl} + H_{\rm ts}
\end{equation}
with 
\begin{adjustwidth}{-0.57em}{}
\begin{subequations}\label{eq-6-2-17}
    \begin{align}
        &
        \begin{aligned}
            H_{\rm bd} \!\!& =\!\!
            {\mathsmaller\sum\nolimits_{i=1}^{T\!-\!1}} \! \big[ \Vert \phi_i^*(\bm{y}\B \!\!-\!\! \bm{y}^{*,i}) \Vert_{2, \bm{w}_{\rm bd}\B }^2 \!\!+\!\! \Vert \phi_i^*(\bm{y}^{*,i} \!\!-\!\! \bm{y}\U) \! \Vert_{2, \bm{w}_{\rm bd}\U }^2 \big]
            \\ 
            &
             +\!\! {\mathsmaller\sum\nolimits_{i=\!1}^{T}} \! \big[ \Vert \phi_i^-\!(\bm{y}\B \!\!-\!\! \bm{y}^{\!-,i}) \Vert_{2, \bm{w}_{\rm bd}\B }^2 \!\!+\!\! \Vert \phi_i^-\!(\bm{y}^{\!-,i} \!\!-\!\! \bm{y}\U) \!\Vert_{2, \bm{w}_{\rm bd}\U}^2 \big] \!\!\!\!\!
        \end{aligned}
        \\
        & 
        \begin{aligned}
            H_{\rm vl} \!= &  {\mathsmaller\sum\nolimits_{i=1}^{T}}\! \big[ \Vert \bm{y}^{*,i-1} \!\!-\! \bm{y}^{-,i} \Vert_{2, \bm{w}_{\rm vl}} \!+\! \Vert \bm{y}^{-,i} \!-\! \bm{y}^{*,i} \Vert_{2, \bm{w}_{\rm vl}} \big]
            \\
            & - \Vert \bm{y}^{*,0} \!-\! \bm{y}^{*,T} \Vert_{2, \bm{w}_{\rm vl}} 
        \end{aligned}
        \\
        & H_{\rm ts} \!=\! {\mathsmaller\sum\nolimits_{i=1}^{T}} \mathsmaller\int_{t_{i}^+}^{t_{i}^*} \Vert \bm{y} - \bm{y}^{*, i} \Vert_{2, \bm{w}_{\rm ts}}^2 \text{d} t
    \end{align}
\end{subequations}
\end{adjustwidth}
where ${H}_{\rm bd}$ and ${H}_{\rm vl}$ represent boundedness and volatility of the SSCs of $\bm{y}$, respectively; ${H}_{\rm ts}$ represents bumpiness of the TSC; $\bm{w}_{\rm bd}\B$, $\bm{w}_{\rm bd}\U$, $\bm{w}_{\rm vl}$, and $\bm{w}_{\rm ts}$ are weight vectors for associated components of $H$; $\bm{y}\B \!\!=\!\! \min( \bm{y}^{*,0}, \bm{y}^{*, T} )$, $\bm{y}\U \!\!=\!\! \max( \bm{y}^{*,0}, \bm{y}^{*, T})$, where $\max(\cdot, \cdot)$ and $\min(\cdot, \cdot)$ are the entry-wise maximum and minimum of the two vectors, respectively; $\phi_i^*(\cdot) = \bm{0}$ if the $i$-th TE contains a fictitious execution of line switching, and $\phi_i^*(\cdot) \!\!=\!\! \max(\cdot, \bm{0})$ otherwise; $\phi_i^-(\cdot)$ is analogous to $\phi_i^*(\cdot)$ but depends on if the $i$-th TE contains a fictitious execution of ACV adjustment.

\subsection{Tractable Surrogates}

In the bumpiness metric, $H_{\rm ts}$ is hard to be incorporated into a mathematical formulation of the BTT problem since it generally relies on discretization transformations or time-domain simulations of (\ref{eq-6-2-13}). These two techniques introduce high dimensionality to the BTT problem with binary variables already, and black-box components, respectively, which cause intractability for solution approaches. Therefore, we develop two surrogates for $H_{\rm ts}$, called $\mathcal{H}_2$-norm surrogate and jumping-state-based surrogate, to ensure tractability of mathematical formulation of the BTT problem.

\subsubsection{$\mathcal{H}_2$-norm surrogate} 

The TSC $\bm{y} \!-\! \bm{y}^{*, i}$ during the $i$-th post-switching phase is the free output response of system $\Psi(\bm{z}^{i}\!,\! \bm{a}^{i})$ to the initial state $(\bm{x}(t_i^+), \bm{\xi}(t_i^+)) \!\!=\!\! (\bm{x}^{+, i}\!, \bm{\xi}^{+, i})$ with a shift $\!- \bm{y}^{*, i}$. 
Denote this TSC by $\Delta \bm{y}_{\rm nl}^i (t)$ with $t \!\!\in\!\! [t_{i}^+\!, t_i^*]$. 
Let $\Delta \bm{y}_{\rm fr}^i(t)$ with $t \!\!\in\!\! [t_{i}^+, t_{i}^*]$ be the free output response of 
$G(s,\! \bm{z}^{i}\!, \bm{a}^{i}\!, \bm{x}^{*, i}\!, \bm{\xi}^{*, i}\!, \cdot)$ to the initial state $\Delta \bm{x}(\!t_{i}^+\!) \!=\! \bm{x}^{+, i} \!-\! \bm{x}^{*, i}$. 
If the jumping state is sufficiently close to the following post-switching SS, $G$ can approximate $\Psi$ regarding the free output response. Formally, we make the following assumption: 
\begin{itemize} 
    \item[A.5] $\forall i \!\in\! \overline{1, T}$, $\Vert \bm{x}^{+, i} \!-\! \bm{x}^{*, i} \Vert_2$ is sufficiently small such that $\Delta \bm{y}_{\rm nl}^i (t) = \Delta \bm{y}_{\rm fr}^i(t) $ with $t \in [t_{i}^+, t_{i}^*]$. 
\end{itemize}

Let $\bm{B}(\bm{x}^{\Delta}) \!=\! (\bm{x}^{\Delta}) \D $, $\bm{u} \!\!=\!\! \bm{1}_{n_{\rm x}} \tilde{\delta}(t \!-\! t_{i}^+) $ 
with $\tilde{\delta}(t)$ being the unit impulse function, and $\Delta \bm{y}_{\rm im}^i(t)$ with $t \!\!\in\!\! [t_i^+, t_{i}^*]$ be the response of $G(s,\bm{z}^{i}, \bm{a}^{i}, \bm{x}^{*, i}, \bm{\xi}^{*, i}, \bm{x}^{+, i} \!-\! \bm{x}^{*, i} )$ with $\bm{x}(t_i^+) \!\!=\!\! \bm{0}$ to the inputs $\bm{u}$. 
Then we have the following proposition:
\begin{proposition}\label{prop-6-2-1}
    $\forall i \!\in\! \overline{1, T}$, $\Delta \bm{y}_{\rm fr}^i (t) \!=\! \Delta \bm{y}_{\rm im}^i(t) $ with $t \!\in\! [t_i^+, t_{i}^*]$.
\end{proposition}

\begin{proof}
    The free output response $\Delta \bm{y}_{\rm fr}^i (t)$ is given by 
    \begin{equation}\label{eq-6-2-19}
        \Delta \bm{y}_{\rm fr}(t) = \bm{C} e^{\bm{A} (t - t_i^+)} (\bm{x}^{+, i} \!-\! \bm{x}^{*, i})  ~~~ t_i^+ \leq t \leq t_{i}^*
    \end{equation}
    The output response $\Delta \bm{y}_{\rm im}^i(t)$ is given by
    \begin{equation}\label{eq-6-2-20}
        \!\!\!\! \Delta \bm{y}_{\rm im}^i\!(t) \!\!=\!\! \bm{C} e^{\bm{A} (t - t_i^+\!)} \!\bm{B}( \bm{x}^{\!+, i} \!\!-\! \bm{x}^{*, i} ) \bm{1} \!\!=\!\! \Delta \bm{y}_{\rm fr}(t) ~~ t_i^+ \!\leq\! t \!\leq\! t_{i}^* \!\!\!
    \end{equation} 
    which completes the proof.
\end{proof}

Under assumption A.5, by Proposition \ref{prop-6-2-1}, we have
\begin{equation}\label{eq-6-2-21}
    \!\!\!\!
    \begin{aligned}
        H_{\rm ts} & \!=\! {\mathsmaller\sum\nolimits_{i=1}^{T}}   \mathsmaller\int_{t_i^+}^{t_{i}^*}  \Vert \Delta \bm{y}_{\rm im}^i(t)  \Vert_{2, \bm{w}_{\rm ts}}^2   \text{d} t \\
        & = {\mathsmaller\sum\nolimits_{i=1}^{T}}  \Vert  \tilde{G}(\!s,\bm{z}^{i}, \bm{a}^{i}, \bm{x}^{*, i}, \bm{\xi}^{*, i}, \bm{x}^{+, i} \!-\! \bm{x}^{*, i} )  \Vert_{\mathcal{H}_2}^2
    \end{aligned} \!\!\!
\end{equation}
where $\tilde{G}$ is the transfer function formed by replacing $\bm{C}$ by $\tilde{\bm{C}} \!\!=\!\! (\bm{w}_{\rm ts}\D)^{\frac{1}{2}} \bm{C}$ in $G$. When $G$ is asymptotically stable, $\mathcal{H}_2$ norm of the transfer function can be computed with the observability Gramian, giving the final tractable $\mathcal{H}_2$-norm surrogate for $H_{\rm ts}$
\begin{equation}\label{6-2-22}
    \!\! H_{\rm ts}'' \!=\!\! {\mathsmaller\sum\nolimits_{i=1}^{T}} \Tr( ( \bm{x}^{+, i} \!-\! \bm{x}^{*, i} )\D \bm{Q}_i ( \bm{x}^{+, i} \!-\! \bm{x}^{*, i} )\D)  
\end{equation}
with $\bm{Q}_i \in \mathbb{R}^{n_{\rm x} \times n_{\rm x}}$ satisfying the Lyapunov equation
\begin{equation}\label{eq-6-2-lya}
    \bm{A}(\bm{z}^{i}, \bm{a}^{i}, \bm{x}^{*, i}, \bm{\xi}^{*, i})\T \bm{Q}_i + \bm{Q}_i \bm{A} = - \bm{C}\T \bm{C}
\end{equation}

\subsubsection{Jumping-state-based surrogate} 
\begin{figure}[h]
	\centering
	\includegraphics[width=\linewidth]{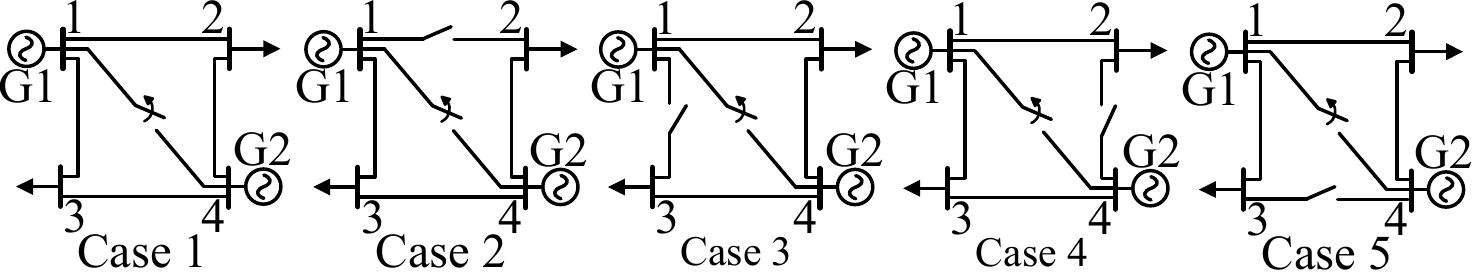} 
	\caption{Illustration of 5 cases of the 4-bus system with different pre-switching network topology.}
	\label{fig-6-2-4} 
\end{figure}

Firstly, we use the 4-bus system as shown in Fig. \ref{fig-6-2-4} to illustrate the observation which inspires this surrogate. In this system, line 1-4 is to be opened, and five cases with different network topology before opening line 1-4 are considered.

\begin{figure}[h]
	\centering
	\includegraphics[width=\linewidth]{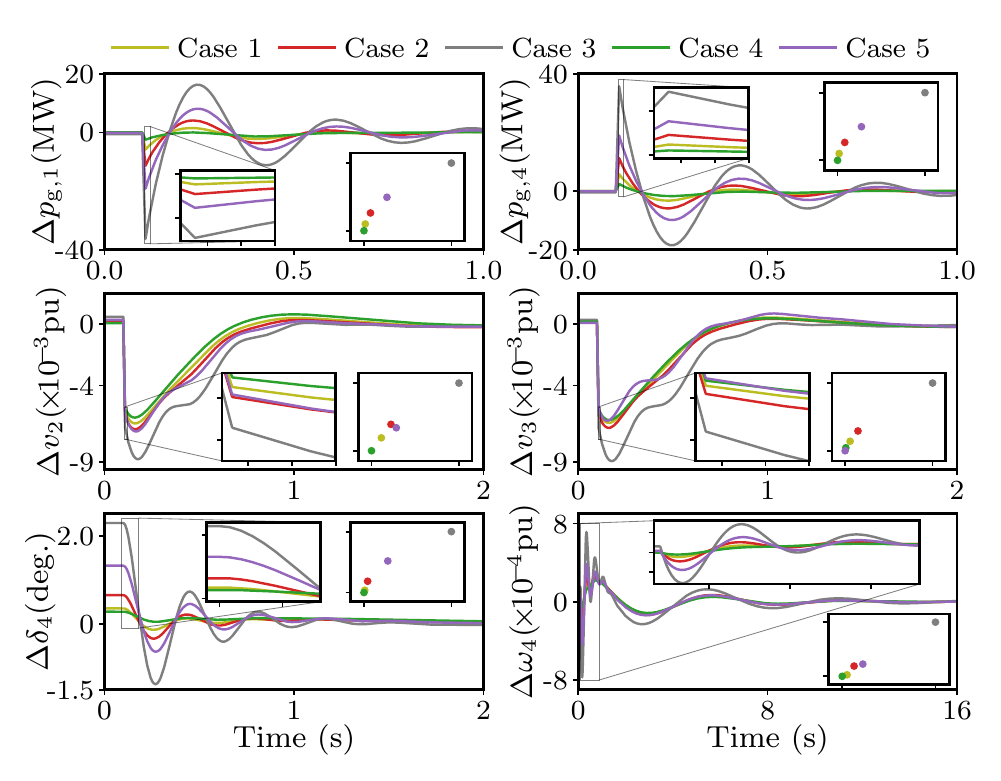} 
	\caption{Post-switching trajectories of the performance outputs for case 1 to 5. The left-side small figure of each subfigure shows the zoom of the trajectories around the jumping state. In the right-side small figure of each subfigure, taking the subfigure for $\Delta p_{\rm g, i}$ as an example, 
    the scatter plot shows the relationship between bumpiness $H_{\rm ts}$ for $\Delta p_{\rm g, 1}$ and the value of $|\Delta p_{\rm g, 1}|$ under the jumping state, and each marker is associated with the case with the same trajectory color. 
    }
	\label{fig-6-2-r1} 
\end{figure}

Fig. \ref{fig-6-2-r1} gives the post-switching trajectories of TSCs of different performance outputs for these cases, and the scatter plots show the relationship between bumpiness and two metrics of the trajectory. The performance outputs here include active power outputs of G1 and G2, voltages at bus 2 and 3, rotor angle of G2 with reference to the rotor angle of G1, and rotor angle speed of G2, whose TSCs are denoted by $\Delta p_{\rm g, 1}$, $\Delta p_{\rm g, 4}$, $\Delta v_2$, $\Delta v_4$, $\Delta \delta_4$, and $\Delta \omega_4$, respectively. 
We can observe that bumpiness of each TSC is proportional to the absolute value of the TSC under the jumping state, except for the bumpiness of $\Delta v_2$ of case 2 and 5. However, $H_{\rm ts}$ for $\Delta v_2$ of case 2 and that for case 5 are very close. Inspired by this observation, minimization of bumpiness of the TSC can be approximately converted into that of the absolute value of the TSC under the jumping state. Accordingly, we propose the jumping-state-based surrogate for $H_{\rm ts}$ given by
\begin{equation}
    H_{\rm ts}' = {\mathsmaller\sum\nolimits_{i=1}^{T}} \Vert \bm{y}^{+, i} - \bm{y}^{*, i} \Vert_{2, \bm{w}_{\rm ts}\D \bm{w}'_{\rm ts} }^2 
\end{equation}
where $\bm{w}'_{\rm ts}$ is the vector of estimated scale factors between $ \mathsmaller\int_{t_i^+}^{t_{i}^*} (\bm{y}_j \!-\! \bm{y}^{*, i}_j)^2  \text{d} t$ and $(\bm{y}^{+, i}_j \!-\! \bm{y}^{*, i}_j)^2$.

\section{Mathematical Formulation of the BTT Problem}

\subsection{High-Level Formulation}

\begin{figure}[h]
	\centering
	\includegraphics[width=8.5cm]{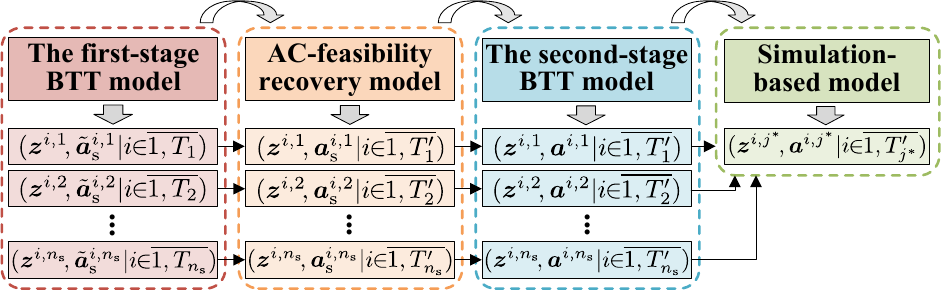} 
	\caption{Illustration of the high-level formulation of the BTT problem.}
	\label{fig-6-2-7} 
\end{figure}

Fig. \ref{fig-6-2-7} gives the high-level formulation of the BTT problem. It contains four submodels as follows.

\subsubsection{The first-stage BTT model} This is an MISOCP model to find the optimal and suboptimal transition trajectories of topology and ACVs in $\bm{a}_{\rm s}$, using the linearized power flow model and the surrogate $H_{\rm ts}'$ for $H_{\rm ts}$. In Fig. \ref{fig-6-2-7}, $n_{\rm s}$ is the number of optimal and suboptimal solutions, 
$(\bm{z}^{i,j}\!, \tilde{\bm{a}}_{\rm s}^{i, j}| i \!\in\! \overline{1,\! T_j})$ with $j \!\!\in\!\! \overline{1,\! n_{\rm s}}$ denotes the $j$-th solution of $(\bm{z}^{i}, \bm{a}_{\rm s}^{i}| i \!\!\in\!\! \overline{1, T} )$ produced by the model and $T_j$ is the value of $T$ for this solution.

\subsubsection{AC-feasibility recovery model} The solutions yielded by the first-stage BTT model may be AC infeasible due to the linearization of power flow models \cite{4-1301}. The model here is an NLP model to find the AC feasible solution of $(\bm{z}^{i}, \bm{a}_{\rm s}^{i}| i \!\!\in\!\! \overline{1, T})$, denoted by $(\!\bm{z}^{i,j}\!\!, \bm{a}_{\rm s}^{i,j}| i \!\!\in\!\! \overline{1,\! T_j'})$ with $T_j'$ being the value of $T$ for it, which is closest to $(\!\bm{z}^{i,j}\!\!, \tilde{\bm{a}}_{\rm s}^{i, j}| i \!\!\in\!\! \overline{1,\! T_j} )$, for $j \!\!\in\!\! \overline{1, n_{\rm s}}$. 

\subsubsection{The second-stage BTT model} Given $(\bm{z}^{i,j}, \bm{a}_{\rm s}^{i,j}| i \!\!\in\!\! \overline{1, T_j'} )$, this model finds the corresponding optimal transition trajectory of ACVs in $\bm{a}_{\rm t}$, denoted by $(\bm{a}_{\rm t}^{i, j} | i \!\!\in\!\! \overline{1, T_j'} )$. Then the transition trajectory of topology and all ACVs is $(\bm{z}^{i,j}\!, \bm{a}^{i, j} | i \!\!\in\!\! \overline{1, T_j'} )$, with $\bm{a}^{i, j} \!=\! [(\bm{a}_{\rm s}^{i, j})\T\!, (\bm{a}_{\rm t}^{i, j})\T]\T$. This model, in an NLP form, uses the $\mathcal{H}_2$-norm surrogate $H_{\rm ts}''$ for $H_{\rm ts}$ and low-fidelity system dynamic models to ensure computational efficiency.

\subsubsection{Simulation-based model} This model, based on high-fidelity  time-domain simulations, gives accurate evaluations of the $n_{\rm s}$ transition trajectories of topology and ACVs, and finds the optimal one, denoted as $(\bm{z}^{i, j^*}\!, \bm{a}^{i, j^*} \!| i \!\!\in\!\! \overline{1,\! T_{\!j^*}'} )$ with $j^* \!\!\in\!\! \overline{1,\! n_{\rm s}}$. Some complicated issues which are ignored before, including stability of the overall transition process and multiple post-switching equilibrium points, are captured by this model.

\subsection{The First-Stage BTT Model}

\subsubsection{Pre-treatment} $\!$The number of TEs $T$ is unknown beforehand. Without assuming any pattern of each TE, 
$T$ is restricted only by (\ref{eq-6-2-a4}). Thus,
the transition trajectory should be modeled with the maximal possible value of $T$ determined by (\ref{eq-6-2-a4}), i.e., 
\begin{equation}
    T\U \!=\! \max\{  \Vert \bm{z}^0  \!\!-\! \bm{z}^T \Vert_1 \!+\! \lfloor\! \frac{T_{\max}}{T_{\rm ls}} \!\rfloor, \Vert \bm{z}^0  \!\!-\! \bm{z}^T \Vert_1 \!+\! \lfloor\! \frac{T_{\max}}{T_{\rm ad}} \!\rfloor \!-\! 1 \}
\end{equation}

\subsubsection{Objective} The objective is to minimize the bumpiness metric with the jumping-state-based surrogate, i.e.,
\begin{equation}\label{eq-6-2-1st-obj}
    \min\nolimits_{(\bm{z}^i, \bm{a}_{\rm s}^i|i \in \overline{1, T\U})}   ~ H' \!=\! H_{\rm bd} + H_{\rm vl} + H'_{\rm ts}
\end{equation}
with $T$ substituted by $T\U$.

\subsubsection{Constraints}

Power flow constraints for the steady and jumping states are
\begin{subequations} 
    \begin{align}
        & \tilde{f}_{\rm p}( \bm{x}_{\rm p}^{-, i} | \bm{z}^{i-1}, \bm{a}_{\rm s}^i | \bm{y}_{\rm p}^{-,i}  ) \leq \bm{0} \\
        & \tilde{f}_{\rm p}( \bm{x}_{\rm p}^{+, i} | \bm{z}^i, \bm{a}_{\rm s}^i | \bm{y}_{\rm p}^{+,i} ) \leq \bm{0} \\
        & \tilde{f}_{\rm p}( \bm{x}_{\rm p}^{*, i} | \bm{z}^i, \bm{a}_{\rm s}^i | \bm{y}_{\rm p}^{*, i} ) \leq \bm{0}
    \end{align}
\end{subequations}
for all $i \in \overline{1, T\U}$. Unless otherwise specified, all constraints in the first-stage BTT model are defined for all $i \in \overline{1, T\U}$.

The ACVs are required to be adjusted to their initial values and topology to $\bm{z}^{T}$ at the end of the transition process. Thus
\begin{equation}\label{eq-6-2-terminal}
    \bm{a}_{\rm s}^{T\U} = \hat{\bm{a}}^0_{\rm s}, \bm{z}^{T\U} = \bm{z}^{T}
\end{equation}
where $\hat{\bm{a}}^0_{\rm s}$ is a given value of $\bm{a}_{\rm s}$. 
If the system with $(\bm{z}, \bm{a}_{\rm s}) \!\!=\!\! (\bm{z}^{T}, \bm{a}_{\rm s}^0)$ is operationally feasible with the linearized power flow, $\hat{\bm{a}}^0_{\rm s} \!\!=\!\! \bm{a}^0_{\rm s}$; and otherwise, $\hat{\bm{a}}^0_{\rm s}$ is the closest one to $\bm{a}^0_{\rm s}$ which makes the system operationally feasible with the linearized power flow. Alternatively, we can also remove the first equality in (\ref{eq-6-2-terminal}), and penalize $-\! \Vert \bm{a}_{\rm s}^{T\U} \!\!\!-\! {\bm{a}}^0_{\rm s} \Vert_2$ in the objective function.

For the post-adjustment and post-switching SSs, values of the ACVs are equal to their setpoints, $\bm{p}_{\rm go}$ are maintained at the initial value, the IM components of load powers also remain constant. Thus we have 
\begin{adjustwidth}{-0em}{}
\begin{subequations}\label{eq-6-2-maintain}
    \begin{align}
        & \bm{a}_{\rm s}^{-, i} = \bm{a}_{\rm s}^i, \bm{p}_{\rm go}^{-, i} = \bm{p}_{\rm go}^{*, 0}, \bm{\epsilon}_{\rm p}^{-,i} = \bm{\epsilon}_{\rm p}^{*,0}, \bm{\epsilon}_{\rm q}^{-,i} = \bm{\epsilon}_{\rm q}^{*,0} \label{eq-6-2-maintain:1} \\
        & \bm{a}_{\rm s}^{*, i} = \bm{a}_{\rm s}^i, \bm{p}_{\rm go}^{*, i} = \bm{p}_{\rm go}^{*, 0}, \bm{\epsilon}_{\rm p}^{*,i} = \bm{\epsilon}_{\rm p}^{*,0}, \bm{\epsilon}_{\rm q}^{*,i} = \bm{\epsilon}_{\rm q}^{*,0}  \label{eq-6-2-maintain:2}
    \end{align}
\end{subequations}
\end{adjustwidth}

For the jumping states, some variables, denoted as a vector $\bm{x}_{\rm c}$, remain the same values as those under the pre-switching SS, which gives
\begin{equation}\label{eq-6-2-inst-same}
    \bm{x}_{\rm c}^{+, i} = \bm{x}_{\rm c}^{-, i} 
\end{equation}
with 
\begin{equation}
    \!\!\!\!\!\!\!  \bm{x}_{\rm c} \!\!=\!\! [\bm{e}\T\!, \bm{\delta}\T\!, \bm{e}_{\rm s}\T\!, \bm{\delta}_{\rm s}\T\!, \bm{v}_{\!\rm m}\T, \bm{\theta}_{\!\rm m}\T, \bm{e}_{\!\rm m}\T, \bm{\delta}_{\!\rm m}\T, \bm{b}_{\!\rm svc}\T, \bm{v}_{\!\rm svg}\T, \bm{\theta}_{\!\rm svg}\T, \bm{b}_{\rm b, tcsc}\T]\T
    \!\!\!\!\!\!\!
\end{equation}

Operational constraints for the SSs include
\begin{subequations}\label{eq-6-2-oprconst}
    \begin{align}
        & \bm{q}_{\rm gs}\B  \!\leq\! \bm{q}_{\rm gs}^{-, i} \!\leq\!   \bm{q}_{\rm gs}\U, [(\bm{p}_{\rm gc}^{-,i})\HP + (\bm{q}_{\rm gc}^{-,i})\HP]\HR \leq \bm{s}_{\rm gc}\U, \langle\! * \!\rangle  \label{eq-6-2-oprconst:1} \\
        & \bm{q}_{\rm gc}^{-, i} \leq \bm{p}_{\rm gc}^{-, i} \circ \tan(\arccos( \bm{\phi}_{\rm gc}\B )), \bm{v}\B \leq \bm{v}^{\!-\!, i} \leq \bm{v}\U, \langle\! * \!\rangle \label{eq-6-2-oprconst:2}\\
        & -\! \bm{\theta}\U \!-\! (\bm{1} \!-\! \bm{z}^{i}) M \!\leq\! \bm{E}\T \!\bm{\theta}^{-, i} \!\!\leq\! \bm{\theta}\U \!+\! (\bm{1} \!-\! \bm{z}^{i}) M, \langle\! * \!\rangle \label{eq-6-2-oprconst:3}\\
        & [ (\bm{p}^{\!-\!, i}_{\rm fb}){\!}\HP \!\!+\!\! (\!\bm{q}^{\!-\!, i}_{\rm fb}){\!}\HP ]\HR \!\!\leq\!\! \bm{s}_{\rm b}\U, 
          [ (\bm{p}^{\!-\!, i}_{\rm tb})\HP \!\!+\!\! (\!\bm{q}^{\!-\!, i}_{\rm tb})\HP ]\HR \!\!\leq\!\! \bm{s}_{\rm b}\U, \langle\! * \!\rangle \label{eq-6-2-oprconst:4} \\
        & \bm{b}_{\rm b, tcsc}\B \!\!\leq\!\! \bm{b}_{\rm b, tcsc}^{-,i} \!\!\leq\!\! \bm{b}_{\rm b, tcsc}\U, 
        \bm{b}_{\!\rm svc}\B \!\!\leq\!\! \bm{b}_{\!\rm svc}^{\!-\!,i} \!\!\leq\!\! \bm{b}_{\!\rm svc}\U, 
        \bm{q}_{\rm c}\B \!\!\leq\!\! \bm{q}_{\rm c}^{\!-\!,i} \!\!\leq\!\! \bm{q}_{\rm c}\U\!, \langle\! * \!\rangle  \label{eq-6-2-oprconst:5}
    \end{align}
\end{subequations}
where (\ref{eq-6-2-oprconst:1}) are output power constraints of generators; (\ref{eq-6-2-oprconst:2}) are constraints for power factors of CIGs and bus voltage magnitudes, respectively; (\ref{eq-6-2-oprconst:3}) and (\ref{eq-6-2-oprconst:4}) bound branch phase angle differences and branch powers, respectively; (\ref{eq-6-2-oprconst:5}) are constraints for equivalent susceptances of lines with TCSC, output susceptances of SVCs, and reactive power outputs of DVCs, respectively; and $\langle\! * \!\rangle$ denotes the same constraints as the left-side but for the post-switching SSs.

The abrupt changes in generator electric power during the switching phase can induce rotor shaft impacts (RSIs), which should be kept within safe levels \cite{4-1285}. The same problem exists for IM loads. Thus we have  
\begin{equation}\label{eq-6-2-rsi}
    |\bm{p}_{\rm gs}^{+, i} - \bm{p}_{\rm gs}^{-, i}|_{\circ} \!\leq\!  \bm{\varepsilon}_{\rm gs}\D   \bm{p}_{\rm gs}^{\scriptscriptstyle\rm N} ~\!\!, 
    \bm{p}_{{\rm{d_0}}}\D | \bm{\epsilon}_{{\rm p}}^{-, i} \!-\! \bm{\epsilon}_{{\rm p}}^{+, i} |_{\circ} \leq 
    \bm{\varepsilon}_{\rm im}\D \bm{p}_{\rm im}^{\scriptscriptstyle\rm N}
\end{equation}
where $\bm{\varepsilon}_{\rm gs}$ and $\bm{\varepsilon}_{\rm im}$ are vectors of the proportions of rated power associated with critical RSI levels, for SGs and IMs, respectively; $\bm{p}_{\rm gs}^{\scriptscriptstyle\rm N}$ and $\bm{p}_{\rm im}^{\scriptscriptstyle\rm N}$ are vectors of rated powers of SGs and IM loads, respectively.

The setpoint of each ACV in $\bm{a}_{\rm s}$ is generally bounded, and the total adjustment amount of ACVs in each TE is also bounded to ensure fast and seamless ACV adjustment. Thus
\begin{equation}\label{eq-6-2-adj-bound}
    \bm{a}_{\rm s}\B \leq \bm{a}_{\rm s}^i \leq \bm{a}_{\rm s}\U, \Vert \bm{a}_{\rm s}^i - \bm{a}_{\rm s}^{i-1} \Vert_{2, \bm{w}_{\rm as} } \leq  \sigma_{\rm as}
\end{equation}
where $\bm{w}_{\rm as}$ is the weight vector, and $\sigma_{\rm as}$ is the maximal allowed total adjustment amount of $\bm{a}_{\rm s}$ in one TE.

For network topology, its connectedness during the transition process should be ensured, thus  
\begin{equation}\label{eq-6-2-connectedness}  
    \!\!\!\! \begin{aligned} 
        & M (\bm{z}^i  \!-\! \bm{1} ) \!\!\leq\!\! \bm{E}^T \!\bm{o}^i \!-\! \bm{\rho}^i  \!\!\leq\!\! M (\bm{1} \!-\! \bm{z}^i )  ~~ i \!\in\! \overline{1, T\U\!\!-\!\!1} \\  
        & \!-\! M \bm{z}^i \!\!\leq\!\! \bm{\rho}^i \!\!\leq\!\! M \bm{z}^i, \bm{E} \bm{\rho}^i \!\!=\!\! \bm{c}, \bm{o}^i \!\!\in\!\! \mathbb{R}^{n_{\rm n}}\!, \bm{\rho}^i \!\!\in\!\! \mathbb{R}^{n_{\rm e}} ~~ i \!\in\! \overline{1, T\U\!\!-\!\!1}
    \end{aligned} 
\end{equation}
with $\bm{c}$ being an $n_{\rm n}$-dimensional constant uniquely-balanced vector (see \cite{4-995-ea} for its definition). 

In addition, at most one line can be switched in each TE by Assumption A.2, and some branches, denoted as $\mathcal{E}_{\rm np}$, do not participate in the auxiliary control for topology transition. Thus we have
\begin{equation}
    \Vert \bm{z}^i \!-\! \bm{z}^{i-1} \Vert_1 \!\leq\! 1, \bm{E}_{\rm np}  (| \bm{z}^0 \!\!-\! \bm{z}^T |_{\circ} \!-\!  {\mathsmaller\sum\nolimits_{i=1}^{T\U}} |\bm{z}^i \!\!-\! \bm{z}^{i-1}|_{\circ}) \!\!=\!\! \bm{0}
\end{equation}
where $\bm{E}_{\rm np}$ is the adjacent matrix between $\mathcal{E}_{\rm np}$ and $\mathcal{E}$.

To formulate constraint (\ref{eq-6-2-a4}), we introduce $\bm{\zeta} \!=\! [\zeta_{i}]$ and $\tilde{\bm{\zeta}} \!\!=\!\! [\tilde{\zeta_{i}}]$, 
with $i \!\!\in\!\! \overline{1, {T\U}}$, $\zeta_{i} \!\!\in\!\! \mathbb{B}$, and $\tilde{\zeta_{i}} \!\!\in\!\! \mathbb{R}$, to indicate the type of each TE. Let $\bm{\zeta}$ and $\tilde{\bm{\zeta}}$ satisfy
\begin{equation}
    \Vert \bm{a}^i_{\rm s} \!-\!  \bm{a}^{i-1}_{\rm s} \Vert_{2, \bm{w}_{\rm as} }  \!\leq\! \sigma_{\!\rm as} \zeta_i, \Vert \bm{z}^{i} - \bm{z}^{i-1} \Vert_1 = \tilde{\zeta}_i
\end{equation}
with $\delta_{\rm pen} \bm{1}\T \bm{\zeta}$ being penalized in the objective function. Here $\delta_{\rm pen} >0 $ is a properly small penalty coefficient to ensure that minimization of $H'$ takes precedence over that of $\bm{1}\T \bm{\zeta}$. \hig{The value of $\delta_{\rm pen}$ can be selected according to the maximum acceptable tolerance of $H'$. For instance, given the maximum acceptable tolerance of $H'$ being $10^{-5}$, $\delta_{\rm pen}$ can be set as a value smaller than $10^{-5}/T_{\rm u}$, such that the decrease of $\delta_{\rm pen} \bm{1}\T \bm{\zeta}$ caused by the minimization of $\bm{1}\T \bm{\zeta}$ is always smaller than the maximum acceptable tolerance of $H'$, i.e., $10^{-5}$.} 
Then, $\zeta_{i}\!=\! 1  \!\Leftrightarrow\!$  ACVs are adjusted in the $i$-th TE, and $\tilde{\zeta_{i}}\!=\! 1  \!\Leftrightarrow\!$  lines are switched in the $i$-th TE. Accordingly, (\ref{eq-6-2-a4}) is equivalent to
\begin{equation}
    T_{\rm ad}  \bm{1}\T \bm{\zeta} + T_{\rm ls} ( \bm{1}\T \tilde{\bm{\zeta}} - \Vert \bm{z}^0  \!\!-\! \bm{z}^T \Vert_1 )   \leq T_{\max}
\end{equation}

Finally, we consider the structure of the sequence of TEs. 
Introduce variables $\bm{\zeta}' \!\!=\!\! [\zeta'_{i}]$ with $i \!\!\in\!\! \overline{1, T\U}$ and $\zeta'_{i} \!\!\in\!\! \mathbb{R}$ satisfying
\begin{equation}
    \bm{\zeta}' \geq \bm{\zeta}, \bm{\zeta}' \geq \tilde{\bm{\zeta}}, \bm{\zeta}' \geq \bm{\zeta} + \tilde{\bm{\zeta}}, \bm{\zeta}' \geq  \bm{1} 
\end{equation}
such that $\zeta'_i = 0 \Leftrightarrow$ the $i$-th TE is with only fictitious executions, and $\zeta'_i = 1 \Leftrightarrow$ otherwise. 
The TEs with only fictitious executions should only appear at the end of the sequence of TEs. This indicates that $\forall i \!\!\in\!\! \overline{1, T\U-2}$, $(\zeta'_i, \zeta'_{i+1}, \zeta'_{i+2}) \!\!\notin\!\! \{(0,1,1), (0,1,0), (0,0,1), (1,0,1) \}$, which is equivalent to
\begin{equation}
    \begin{bmatrix}
        1 & 1 & -1 \\
        1 & -1 & 1\\
        1 & -1 & -1\\
        -1 & 1 & -1
    \end{bmatrix}
    \begin{bmatrix}
        \zeta'_i \\
        \zeta'_{i+1} \\
        \zeta'_{i+2}
    \end{bmatrix}
    \geq 
    \begin{bmatrix}
        0 \\
        0 \\
        -1 \\
        -1
    \end{bmatrix}
    ~~\forall i \!\!\in\!\! \overline{1, T\U-2}
\end{equation}
Additionally, for any adjacent TEs, the former one with only execution of ACV adjustment followed by another with only line switching is not allowed since they should be combined into one TE. This indicates that $\forall i \!\!\in\!\! \overline{1, T\U\!\!-\!\!1}, (\zeta_i,\! \tilde{\zeta}_i,\! \zeta_{i\!+\!1},\! \tilde{\zeta}_{i\!+\!1}) \!\!\neq\!\! (1,0,0,1)$, which can be ensured by
\begin{equation}
    \zeta_i  + \tilde{\zeta}_{i+1} - (\tilde{\zeta}_i  + \zeta_{i+1}) \leq 1 ~~ \forall i \in \overline{1, T\U-1}
\end{equation}

\subsubsection{Post-treatment} 

For the any $j$-th optimal solution given by the above optimization model, removing all its invalid TEs whose associated $\zeta'_i$ equals to 0 yields the associated solution of the first-stage BTT model, i.e., $(\bm{z}^{i,j}, \tilde{\bm{a}}_{\rm s}^{i, j} | i \in \overline{1, T_j})$.

\subsection{AC-Feasibility Recovery Model}

\begin{figure}[h]
	\centering
	\includegraphics[width=\linewidth]{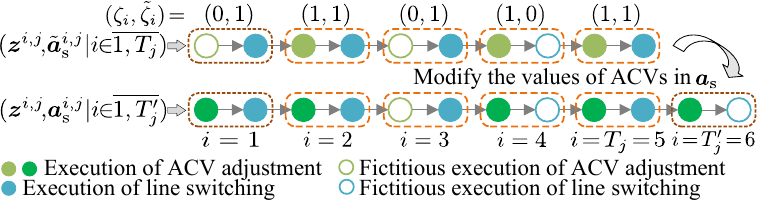} 
    \vspace*{-4pt}
	\caption{Illustration of the pre-treatment and post-treatment.
    }  
	\label{fig-6-2-8} 
\end{figure}

\subsubsection{Pre-treatment}

As shown in Fig. \ref{fig-6-2-8}, the AC-feasibility of $(\bm{z}^{i,j}\!, \tilde{\bm{a}}_{\rm s}^{i, j} | i \!\!\in\!\! \overline{1, T_j})$ is recovered by altering the values of ACVs in $\bm{a}_{\rm s}$ for each execution of ACV adjustment. The structure of the transition process and the topology transition trajectory stay unchanged, except for two cases. The first case is that the first TE only contains line switching, for which a potential execution of ACV adjustment is added to the first TE. The second case is related to the solution of the AC-feasibility recovery model, which is explained in the later post-treatment.

Further by the values of $\bm{\zeta}$ and $\tilde{\bm{\zeta}}$ corresponding to the $j$-th solution, we divide $\overline{1, T_j}$ into three subsets, i.e., $\mathbb{T}_{\rm as}^j$ containing TEs with both ACV adjustment and line switching, $\mathbb{T}_{\rm a}^j$ containing TEs with only ACV adjustment, and $\mathbb{T}_{\rm s}^j$ containing TEs with only line switching. For example, for the $j$-th solution in Fig. \ref{fig-6-2-8}, we have $\mathbb{T}_{\rm as}^j \!=\! \{1,2,5\}$, $\mathbb{T}_{\rm a}^j \!=\! \{4\}$, and $\mathbb{T}_{\rm s}^j \!=\! \{3\}$. Let $\mathbb{T}_{\rm aa}^j =  \mathbb{T}_{\rm as}^j \cup \mathbb{T}_{\rm a}^j$ and $\mathbb{T}_{\rm ss}^j =  \mathbb{T}_{\rm as}^j \cup \mathbb{T}_{\rm s}^j$.

\subsubsection{Formulation}

The AC-feasibility recovery model is formulated as 
\begin{adjustwidth}{-0.57em}{}
\begin{subequations}\label{eq-6-2-recovery-model}
    \begin{align}
        &\!\!\!\!\!\!\!\!\!\!\! \min\nolimits_{ (\bm{a}_{\rm s}^i | i \in \overline{1, T_j})}   \!\mathsmaller\sum\nolimits_{i\in \mathbb{T}_{\rm aa}^j} \! w_{{\rm ac}, i}  \Vert  \bm{a}_{\rm s}^i \!\!-\! \tilde{\bm{a}}_{\rm s}^{i, j}  \!\Vert_{2} 
        \!\!+\!  w'_{\!\rm ac} \Vert  \bm{a}_{\rm s}^{T_j} \!\!-\! {\bm{a}}_{\rm s}^0  \Vert_{2}   \\
        \text{s.t.}~ &  {f}_{\rm p}( \bm{x}_{\rm p}^{-, i} | \bm{z}^{i-1, j}, \bm{a}_{\rm s}^i | \bm{y}_{\rm p}^{-,i}  ) \!\leq\! \bm{0}  ~~ \forall i \!\in\! \mathbb{T}_{\rm aa}^j   \\
        & {f}_{\rm p}( \bm{x}_{\rm p}^{+, i} | \bm{z}^{i, j}, \bm{a}_{\rm s}^i | \bm{y}_{\rm p}^{+,i} ) \!\leq\! \bm{0} ~~ \forall i \!\in\! \mathbb{T}_{\rm ss}^j  \\
        & {f}_{\rm p}( \bm{x}_{\rm p}^{*, i} | \bm{z}^{i, j}, \bm{a}_{\rm s}^i | \bm{y}_{\rm p}^{*, i} ) \!\leq\! \bm{0} ~~\forall  i \!\in\! \mathbb{T}_{\rm ss}^j  \\
        & \{ \text{(\ref{eq-6-2-maintain:1})}, \text{(\ref{eq-6-2-oprconst}) excl.~} \langle\! * \!\rangle, \text{(\ref{eq-6-2-adj-bound})} \} ~~ \forall i \!\in\! \mathbb{T}_{\rm aa}^j    \\
        & \{\text{(\ref{eq-6-2-maintain:2})}, \langle\! * \!\rangle ~\text{in (\ref{eq-6-2-oprconst})}\}  ~~ \forall i \!\in\! \mathbb{T}_{\rm ss}^j  \\
        & \bm{a}_{\rm s}^i \!=\! \bm{a}_{\rm s}^{i-1}  ~~ \forall i \!\in\! \mathbb{T}_{\rm s}^j; 
        \{\text{(\ref{eq-6-2-inst-same}), (\ref{eq-6-2-rsi})} \} ~~ \forall i \!\in\! \mathbb{T}_{\rm as}^j \\
        & \begin{bmatrix}
            \bm{x}_{\rm c}^{+, i} = \bm{x}_{\rm c}^{-, i-1} \\
             |\bm{p}_{\rm gs}^{+, i} \!-\! \bm{p}_{\rm gs}^{-, i-1}|_{\circ} \!\leq\!  \bm{\varepsilon}_{\rm gs}\D   \bm{p}_{\rm gs}^{\scriptscriptstyle\rm N} \\
            \bm{p}_{{\rm{d_0}}}\D | \bm{\epsilon}_{{\rm p}}^{-, i-1} \!-\! \bm{\epsilon}_{{\rm p}}^{+, i} |_{\circ} \!\leq\! \bm{\varepsilon}_{\rm im}\D \bm{p}_{\rm im}^{\scriptscriptstyle\rm N}  
        \end{bmatrix}  ~ \forall (i, i\!-\!\!1) \!\!\in\! \mathbb{T}_{\rm s}^j \!\!\times\!\! \mathbb{T}_{\rm a}^j  \\
        & \begin{bmatrix}
            \bm{x}_{\rm c}^{+, i} = \bm{x}_{\rm c}^{*, i-1} \\
            |\bm{p}_{\rm gs}^{+, i} \!-\! \bm{p}_{\rm gs}^{*, i-1}|_{\circ} \!\leq\!  \bm{\varepsilon}_{\rm gs}\D   \bm{p}_{\rm gs}^{\scriptscriptstyle\rm N} \\
            \bm{p}_{{\rm{d_0}}}\D | \bm{\epsilon}_{{\rm p}}^{*, i-1} \!-\! \bm{\epsilon}_{{\rm p}}^{+, i} |_{\circ} \!\leq\! \bm{\varepsilon}_{\rm im}\D \bm{p}_{\rm im}^{\scriptscriptstyle\rm N}  
        \end{bmatrix}  ~ \forall (i, i\!-\!1) \!\in\! \mathbb{T}_{\rm s}^j  \!\!\times\!\! \mathbb{T}_{\rm ss}^j 
    \end{align}
\end{subequations}
\end{adjustwidth}
where if $i\!=\!1$ and its associated ACV adjustment is added in the pre-treatment, $w_{{\rm ac}, i} \!\!\gg\!\! 1$, and otherwise, $w_{{\rm ac}, i} \!=\! 1$; $w'_{\rm ac} \gg 1$.

\subsubsection{Post-treatment}

Let $(\bm{a}_{\rm s}^{i, j} | i \!\!\in\!\! \overline{1, T_j})$ be the optimal solution of (\ref{eq-6-2-recovery-model}). When $\bm{a}_{\rm s}^{T_j, j} \!\neq\! \bm{a}_{\rm s}^0$, as illustrated in Fig. \ref{fig-6-2-8}, we add a TE with only execution of ACV adjustment following the $T_j$-th TE, which adjusts $\bm{a}_{\rm s}$ from $\bm{a}_{\rm s}^{T_j, j}$ to $\bm{a}_{\rm s}^{T_j + 1, j} = \bm{a}_{\rm s}^{0}$. Let $\bm{z}^{T_j + 1,j} = \bm{z}^{T_j,j}$, and $T_j' \!=\! T_j +\! 1$ if $\bm{a}_{\rm s}^{T_j, j} \neq \bm{a}_{\rm s}^0$ and $T_j' \!=\! T_j$ otherwise. Then the AC-feasible solution corresponding to $(\bm{z}^{i,j}, \tilde{\bm{a}}_{\rm s}^{i, j} | i \in \overline{1, T_j})$ is $(\bm{z}^{i,j}, \bm{a}_{\rm s}^{i, j} | i \in \overline{1, T_j'})$.

\subsection{The Second-Stage BTT Model}

For each $i \!\!\in\!\! \mathbb{T}_{\rm ss}^j$, denote the values of $\bm{x}_{\rm p}^{\!+, i}$ and $\bm{y}_{\rm p}^{\!+, i}$ associated with the AC-feasible solution $(\bm{z}^{i,j}\!, \bm{a}_{\rm s}^{i, j} | i \!\!\in\!\! \overline{1, T_j'})$ as $\bm{x}_{\rm p}^{+, i, j}$ and $\bm{y}_{\rm p}^{+, i, j}$, respectively. They are by-products of solving (\ref{eq-6-2-recovery-model}). Then the jumping state associated the $i$-th TE and $j$-th solution, denoted as $\bm{x}^{+, i, j}$, can be obtained by solving
\begin{equation}
    \begin{aligned}
        & \bm{0} = f_{\rm g}( \bm{x}^{+, i, j}, \bm{\xi}^{+, i, j}, \bm{z}^{i, j}, \cdot )  \\
        & \Xi(\bm{x}^{+, i, j}, \bm{\xi}^{+, i, j}) =  \Xi_{\rm p} (\bm{x}_{\rm p}^{\!+, i, j}, \bm{y}_{\rm p}^{\!+, i, j}) 
    \end{aligned}
\end{equation}
where $\Xi(\bm{x}^{+}\!, \bm{\xi}^{+}\!)$ and $\Xi_{\rm p}(\bm{x}_{\rm p}^{\!+}, \bm{y}_{\rm p}^{\!+})$ return the shared variables among $(\bm{x}^{+}, \bm{\xi}^{+})$ and $(\bm{x}_{\rm p}^{\!+}, \bm{y}_{\rm p}^{\!+})$; $f_{\rm g}(\cdot)$ is formed by removing components of $f(\cdot)$ and $g(\cdot)$ in (\ref{eq-6-2-13}) which only depend on $\Xi(\bm{x}^{+}, \bm{\xi}^{+})$. Analogously, the pre-switching SS, denoted as $\bm{x}^{*, i, j}$ can be yielded.

Then, the second-stage BTT model is formulated as 
\begin{subequations}\label{eq-6-2-second-stage-model}
    \begin{align}
        & \!\!\!\!\!\!\!\!\!\!\!\!\!
        \min_{(\bm{a}_{\rm t}^i | i \in \overline{1, T_j'})}   \mathsmaller\sum\limits_{i \in \mathbb{T}_{\rm ss}^j } \!\! \Tr( ( \bm{x}^{+, i, j} \!\!-\! \bm{x}^{*, i, j} )\D \bm{Q}_i ( \bm{x}^{+, i, j} \!\!-\! \bm{x}^{*, i, j} )\D) 
        \label{eq-6-2-second-stage-model:1} \\
         \text{s.t.} ~&  
         \begin{bmatrix}
            \! \bm{A}(\!\bm{z}^{i, j}\!, [(\!\bm{a}_{\rm s}^{i, j}){\!}\T \!, (\!\bm{a}_{\rm t}^i){\!}\T]\T\!, \bm{x}^{*, i, j}\!, \bm{\xi}^{*, i, j})\T \bm{Q}_i  \\
            + \bm{Q}_i \bm{A} = - \bm{C}\T \bm{C} 
         \end{bmatrix}  \forall i \!\in\! \mathbb{T}_{\rm ss}^j \label{eq-6-2-second-stage-model:2} \\
         & 
         \bm{a}_{\rm t}\B \leq \bm{a}_{\rm t}^i \leq \bm{a}_{\rm t}\U ~~ \forall i \!\in\! \mathbb{T}_{\rm ss}^j; 
         \bm{a}_{\rm t}^{i} = \bm{a}_{\rm t}^{i-1}  ~ \forall i \in \mathbb{T}_{\rm a}^j \label{eq-6-2-second-stage-model:3} \\
         & \bm{Q}_{i} \succ 0, \bm{Q}_{i} \in \mathbb{R}^{n_{\rm x} \times n_{\rm x}} ~~ \forall i \!\in\! \mathbb{T}_{\rm ss}^j  \label{eq-6-2-second-stage-model:4}
    \end{align}
\end{subequations}
where the objective function is $H_{\rm ts}''$ with zero terms removed; 
(\ref{eq-6-2-second-stage-model:2}) are the Lyapunov equations as (\ref{eq-6-2-lya}); 
(\ref{eq-6-2-second-stage-model:3}) are bound constraints for $\bm{a}_{\rm t}$ and equality constraint for $\bm{a}_{\rm t}$ in TEs without line switching; and in (\ref{eq-6-2-second-stage-model:4}), $\bm{Q}_i \succ 0$ ensures asymptotically stability of $G$. Solving (\ref{eq-6-2-second-stage-model})  yields $(\bm{a}_{\rm t}^{i, j} | i \in \overline{1, T_j'})$.

\subsection{Simulation-Based Model}

The simulation-based model yields the final optimal solution with its index $j^*$ given by
\begin{adjustwidth}{-0.57em}{}
\begin{subequations}\label{eq-6-2-sim-1}
    \begin{align}
        j^* = \argmin\nolimits_{j \in \overline{1, n_{\rm s}}} ~&  H((\bm{a}^{i, j}, \bm{z}^{i, j} | i \!\in\! \overline{1, T_j'}))  \\
        \text{s.t.}  ~& \text{The transition process is stable}  \label{eq-6-2-sim-2}\\
        & \{\text{(\ref{eq-6-2-oprconst}), (\ref{eq-6-2-rsi})} \} ~~ \forall i \in \overline{1, T_j'}
        \end{align}
\end{subequations}
\end{adjustwidth}
where the objective function and constraints are all evaluated by high-fidelity time-domain simulations; \hig{constraint (\ref{eq-6-2-sim-2}) ensures stability of the system during the transition process corresponding to the final optimal solution; and the existence of feasible solutions of (\ref{eq-6-2-sim-1}) can be ensured by increasing $n_{\rm s}$.}

\begin{remark}
    \hig{Tripping of a highly loaded transmission line may cause transient instability of the system. 
    The BTT formulation can avoid such tripping although no explicit associated constraints are contained. Specifically, in the first-stage BTT model, minimization of $H'_{\rm ts}$ in (\ref*{eq-6-2-1st-obj}) can avoid switching off a highly loaded line as much as possible since this will generally cause large changes of the performance outputs from the pre-switching state to the jumping state, and thus a large value of $H'_{\rm ts}$. Even if some of the $n_{\rm s}$ solutions yielded by the first-stage BTT model finally cause instability due to tripping of a highly loaded transmission line, the simulation-based model will not select them as the final optimal solution.}
\end{remark}

\section{Numerical Examples}

\hig{This section numerically studies the proposed methodology of BTT. We employ the modified IEEE 9-bus system to intuitively show the basic mechanism and effectiveness of the proposed methodology of BTT. Then, we test for the modified IEEE 118-bus system to demonstrate the effectiveness on realistic-scale networks under various transition scenarios. These two systems both contain CIGs and ESSs for renewable energy. Detailed data of the test cases can be found in \cite{4-971-3}. PowerFactory 2021 is used to solve the simulation-based model on a Windows 10 64-Bit PC with an Intel(R) Core(TM) i5-6500 CPU@3.20GHz and 16GB RAM; Gurobi 9.1 is used to solve the first-stage BTT model, IPOPT 3.14 the other two models, all on a Linux 64-Bit server with 2 Intel(R) Xeon(R) E5-2640 v4 @ 2.40GHz CPUs and 125GB RAM.}
\hig{
With $n_{\rm s}$ set as 4, to find the 4 optimal and suboptimal solutions of the first-stage BTT model, we set the Gurobi parameter ``PoolSearchMode'' to 1 and ``SolutionNumber'' to $4$, which causes Gurobi to search for 1 optimal and $3$ suboptimal solutions. The optimality gap of Gurobi is set to 0.5\%.}

\subsection{The Modified IEEE 9-Bus System}

\begin{figure}[h]
	\centering
	\includegraphics[width=\linewidth]{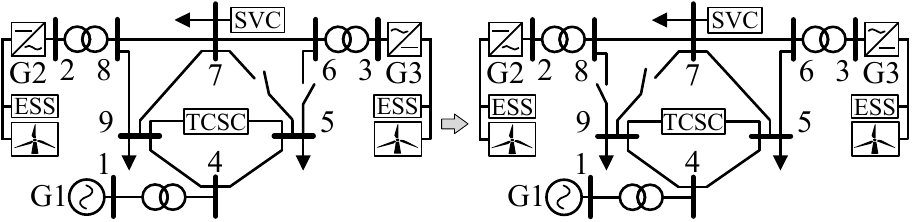} 
	\caption{Diagram of the modified IEEE 9-bus system.
    }  
	\label{fig-6-2-9} 
\end{figure}

The diagram of the modified IEEE 9-bus system is shown in Fig. \ref{fig-6-2-9}. This system contains 1 SG, 2 CIGs both with ESSs, a SVC at bus 7, and a TCSC at line 5-9. The initial and final topology correspond to the left- and right-side network in Fig. \ref{fig-6-2-9}, respectively. 
\hig{The ACVs $\bm{a}$ consist of $v_1$, $v_2$, $v_3$, $p_{\rm g, 2}$,  $p_{\rm g, 3}$, $v_7$, $b_{\rm b, 5\text{-}9}$, $m_{\rm cg, 2}$, $m_{\rm cg, 3}$, $d_{\rm cg, 2}$, and $d_{\rm cg, 3}$}. For a clear illustration, only two performance outputs are considered, i.e., the voltage magnitude at bus 5 denoted as $v_5$ and modulation voltage angle of G2 with reference to the rotor angle of G1 denoted as $\tilde{\theta}_{\rm m, 2}$, so that $\bm{y} = [v_5, \tilde{\theta}_{\rm m, 2}]\T$. 

The optimal transition scheme yielded by the proposed BTT methodology is shown in Table \ref{tab-6-2-2}. \hig{This transition scheme contains 5 TEs among which the first four are complete TEs contianing both ACV adjustment and line switching and the 5th TE contains only ACV adjustment. In the 1st TE, all ACVs are together adjusted from the initial values to the target values smoothly. Taking ACV $v_1$ for example, it is adjusted from the initial value 1.04 to 1.0782. After the system reaches a SS, line 8-9 is switched off. When the system reaches the post-switching SS, the 2ed TE is performed similarly. In the last TE, adjusting all ACVs to their initial values completes the topology transition. For example, $v_1$ is adjusted from 1.0434 to its initial value 1.04 in the 5th TE.}

For comparison purposes, we consider the following four transition schemes:
\begin{itemize}
    \item S1: the optimal transition scheme yielded by the proposed BTT methodology;
    \item S2: the feasible transition schemes without ACV adjustment and minimizing $H$;
    \item S3: the transition scheme given by the optimal topology transition model in \cite{4-1309} which ignores the transients during topology transition processes and ACV adjustment;
    \item S4: the feasible transition schemes without ACV adjustment and maximizing $H$.  
\end{itemize}
Here S2 and S4 can be obtained by solving the proposed BTT model with corresponding modifications. Table \ref{tab-6-2-3} shows the transition schemes S2 to S4, and Table \ref{tab-6-2-4} gives the values of $H$, $H_{\rm bd}$, $H_{\rm vl}$ and $H_{\rm ts}$ associated with the four transition schemes. 

\begin{table}[h]
    \caption{The optimal transition scheme S1}\label{tab-6-2-2}
    \centering
    \setlength{\tabcolsep}{2.5pt} 
    \small{
    \begin{tabularx}{1.0\linewidth}{p{0.5cm}p{1cm}<{\centering}p{1cm}<{\centering}p{1cm}<{\centering}p{1cm}<{\centering}p{1cm}<{\centering}p{1cm}<{\centering}p{1cm}<{\centering}}
    \hline\hline   
    TE & $v_1$  & $v_2$  & $v_3$  & $p_{\rm g, 2}$ &  $p_{\rm g, 3}$ & $v_7$  & $b_{\rm b, 5\text{-}9}$\\ \hline
    0  & 1.0400 & 1.0250 & 1.0250 &  163.00        &  85.00          & 1.0250 & -9.050  \\
    1  & 1.0782 & 1.0231 & 0.9575 &  154.52        &  93.48          & 1.0156 & -39.65  \\
    2  & 1.0790 & 1.0236 & 0.9614 &  158.97        &  89.03          & 1.0119 & -25.21  \\
    3  & 1.0644 & 1.0171 & 0.9636 &  153.28        &  94.72          & 1.0321 & -39.65  \\
    4  & 1.0434 & 1.0090 & 1.0175 &  149.30        &  98.70          & 1.0367 & -6.511  \\
    5  & 1.0400 & 1.0250 & 1.0250 &  163.00        &  85.00          & 1.0250 & -9.050  \\  
    \end{tabularx}
    \begin{tabularx}{1.0\linewidth}{p{0.5cm}p{1cm}<{\centering}p{1cm}<{\centering}p{1cm}<{\centering}p{1cm}<{\centering}p{1.5cm}<{\centering}p{1.5cm}<{\centering}}
        \hline\hline 
    TE & $m_{\rm cg, 2}$ & $m_{\rm cg, 3}$ & $d_{\rm cg, 2}$ & $d_{\rm cg, 3}$ &  Close line & Open line \\ \hline
    0  &  3.0000         &  4.0000         & 10.000          &  20.000         & ---  & --- \\
    1  &  0.4244         &  0.6056         & 7.2692          &  30.000         & ---  & 8-9  \\
    2  &  0.3758         &  0.7332         & 9.1833          &  24.638         & 5-6  & ---  \\
    3  &  0.3499         &  0.3741         & 6.2941          &  12.764         & 5-7  & ---  \\
    4  &  1.1746         &  0.8563         & 13.539          &  30.000         & ---  & 7-9  \\
    5  &  3.0000         &  4.0000         & 10.000          &  20.000         & ---  & ---  \\ 
    \hline\hline
    \end{tabularx}
    }
\end{table}

\vspace*{-10pt}
 
\begin{table}[h]
    \caption{Transition schemes S2, S3 and S4}\label{tab-6-2-3}
    \centering
    \setlength{\tabcolsep}{1.6pt} 
     
    \small{
    \begin{tabularx}{1.0\linewidth}{p{0.25cm}p{1.3cm}<{\centering}p{1.3cm}<{\centering}p{1.3cm}<{\centering}p{1.3cm}<{\centering}p{1.3cm}<{\centering}p{1.3cm}<{\centering}}
    \hline\hline   
    \multirow{2}{*}{$\!\!$TE} & \multicolumn{2}{c}{S2}  & \multicolumn{2}{c}{S3} & \multicolumn{2}{c}{S4} \\ 
    \cmidrule(l){2-3}\cmidrule(l){4-5}\cmidrule(l){6-7} 
      &  Close line & Open line & Close line & Open line & Close line & Open line   \\ \hline
    1  &  --- & 8-9 &  5-7 & ---  &  --- & 7-9   \\ 
    2  & 5-6  & --- &  --- & 7-9  &  5-6 & ---   \\
    3  & 5-7  & --- &  --- & 8-9  &  --- & 8-9   \\
    4  & ---  & 7-9 &  5-6 & ---  &  5-7 & ---   \\  \hline\hline
    \end{tabularx}
    }
\end{table}

\begin{table}[h]
    \caption{Values of the bumpiness metrics for schemes S1 to S4}\label{tab-6-2-4}
    \centering
    \setlength{\tabcolsep}{1.6pt} 
     
    \small{
    \begin{tabularx}{0.7\linewidth}{p{1cm}<{\centering}p{1.2cm}<{\centering}p{1.2cm}<{\centering}p{1.2cm}<{\centering}p{1.2cm}<{\centering}}
    \hline\hline   
    Scheme &   $H$  & $H_{\rm bd}$ & $H_{\rm vl}$ & $H_{\rm ts}$  \\ \hline
    S1  & 0.55 & 0.000895 &  0.148 & 0.397       \\ 
    S2  & 1.11 & 0.0299   &  0.174 & 0.908    \\
    S3  & 1.14 & 0.00959  &  0.160 & 0.973     \\ 
    S4  & 10.6 & 0.3977   &  0.653 & 9.51    \\  \hline\hline
    \end{tabularx}
    }
\end{table}

According to the results in Table \ref{tab-6-2-4}, we have the following observations and conclusions:

\textit{(1)} It is seen that the values of the bumpiness metric and its components of S4 are much larger than that of the other transition schemes. This indicates that a transition scheme with only operational feasibility ensured can significantly underperform regarding bumpiness, which shows the necessity of BTT mechanisms. 

\textit{(2)} The values of $H_{\rm bd}$ and $H_{\rm vl}$ of S3 are smaller than that of S2, while the values of $H_{\rm ts}$ and $H$ are the opposite. Hence, ignoring the transients during topology transition processes as in \cite{4-1309} can result in suboptimal transient bumpiness and thus suboptimal overall bumpiness. 

\textit{(3)} Transition scheme S1 outperforms the other transition schemes in all aspects of bumpiness. Specifically, we can observe more than 90\% improvement in boundedness and 50\% in transient bumpiness by S1 compared with the best of the other schemes. Therefore, by optimally adjusting ACVs during the topology transition process, the proposed BTT methodology achieves much more bumpless topology transition.

\begin{figure}[h]
	\centering
	\includegraphics[width=\linewidth]{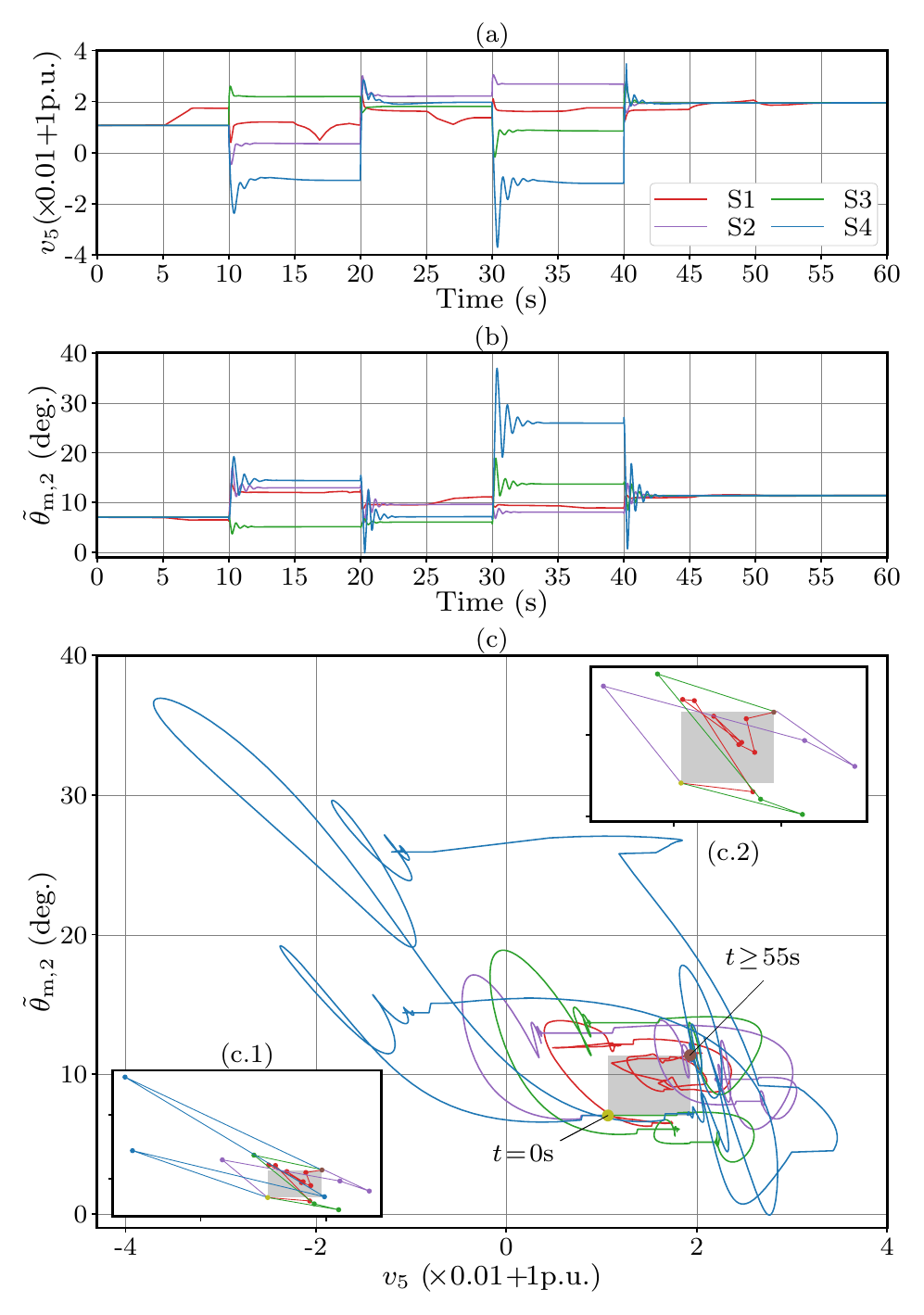} 
    \vspace{-4pt}
	\caption{
     Trajectories of $v_5$-$t$ (a), $\tilde{\theta}_{\rm m, 2}$-$t$ (b) and $v_5$-$\tilde{\theta}_{\rm m, 2}$ (c) during a topology transition with different schemes, and the SSCs of the trajectories of $v_5$-$\tilde{\theta}_{\rm m, 2}$ (c.1 and c.2). Each dot represents a SS. \hig{(c.1) is for schemes S1 to S4, and (c.2) is for only schemes S1 to S3 for better clarity.}
    }  
	\label{fig-6-2-r2} 
\end{figure}

Intuitively, Fig. \ref{fig-6-2-r2} shows the trajectories of the performance outputs during the transition process with the 4 schemes, and the associated SSCs. 
\hig{According to Fig. \ref{fig-6-2-r2} and Table \ref*{tab-6-2-2}, with the optimal transition scheme S1, in the 1st TE, ACVs start to be adjusted at $t=5$s and reach the target values at $t=7$s. The performance outputs $v_5$ and $\tilde{\theta}_{\rm m, 2}$ are changed during this period and reach steady states before $t=10$s. At $t=10$s, line 8-9 is switched off, inducing fast changes of $v_5$ and $\tilde{\theta}_{\rm m, 2}$. The next TE is performed similarly at $t = 15$s where $v_5$ and $\tilde{\theta}_{\rm m, 2}$ reach steady states. The entire transition process is finished at about $t = 55$s where the system reaches the steady state after ACVs are adjusted to their initial values in the 5th TE.} 

Additionally, in Fig. \ref{fig-6-2-r2}-(c), it is seen that the trajectory of $v_5$-$\tilde{\theta}_{\rm m, 2}$ and its SSC for S1 are both bounded better by the optimal region (i.e., the gray rectangle in Fig. \ref{fig-6-2-r2}-(c)) or more close to it compared with that for the other schemes. For the TSCs of $v_5$ or $\tilde{\theta}_{\rm m, 2}$, by Fig. \ref{fig-6-2-r2}-(a) and Fig. \ref{fig-6-2-r2}-(b), obvious post-switching oscillations can be seen for S2 to S4. In contrast, the topology transition with S1 causes no oscillations almost and the smallest overshoots of the TSCs in general.

\subsection{\hig{The Modified IEEE 118-Bus System}}

\hig{
Based on the dynamic IEEE 118-bus system developed in \cite{4-1382}, the modified version utilized in this work replaces 10 SGs by CIGs with ESSs, and 10 SGs working as compensators by 8 SVCs and 2 STATCOMs; models all loads as ZIP-IM loads; and installs 6 TCSCs. The ACVs $\bm{a}$ consist of all 90 eligible variables. The performance outputs $\bm{y}$ consist of voltage magnitudes of all load buses; modulation voltage angles of all CIGs, and rotor angles of all SGs, all with reference to the rotor angle of the reference machine. We consider 200 topology transition scenarios with different initial and final topologies $\bm{z}^0$ and $\bm{z}^T$. For each scenario, we obtain transition schemes S1 to S4 similarly to the test for the modified IEEE 9-bus system, and compute the following ratios: 
\begin{itemize}
    \item $\rho_{4*, \rm{h}}$ ($\rho_{4*, \rm{bd}}$, $\rho_{4*, \rm{vl}}$ and $\rho_{4*, \rm{ts}}$): the ratio of $H$ ($H_{\rm bd}$, $H_{\rm vl}$, and $H_{\rm ts}$) of transition scheme S4 to the largest value of $H$ ($H_{\rm bd}$, $H_{\rm vl}$, and $H_{\rm ts}$) of transition schemes S1 to S3;   
    \item $\rho_{23, \rm{h}}$ ($\rho_{23, \rm{bd}}$, $\rho_{23, \rm{vl}}$ and $\rho_{23, \rm{ts}}$): the ratio of $H$ ($H_{\rm bd}$, $H_{\rm vl}$, and $H_{\rm ts}$) of transition scheme S2 to that of S3;
    \item $\rho_{1*, \rm{h}}$ ($\rho_{1*, \rm{bd}}$, $\rho_{1*, \rm{vl}}$ and $\rho_{1*, \rm{ts}}$): the ratio of $H$ ($H_{\rm bd}$, $H_{\rm vl}$, and $H_{\rm ts}$) of transition scheme S1 to the smallest value of $H$ ($H_{\rm bd}$, $H_{\rm vl}$, and $H_{\rm ts}$) of transition schemes S2 to S4.
\end{itemize}
}

\begin{figure}[h]
	\centering
	\includegraphics[width=\linewidth]{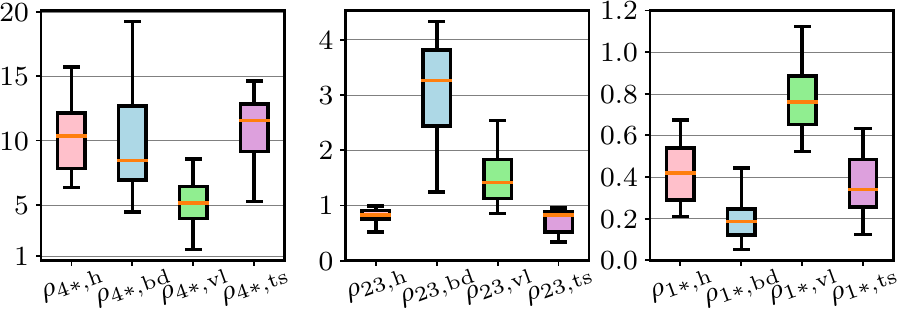} 
	\caption{
     \hig{Boxplots of the ratios for the 500 different transition scenarios.}
    }  
	\label{fig-6-2-r3} 
\end{figure}

\hig{
Fig. \ref*{fig-6-2-r3} shows the boxplots of the above ratios for the 200 different transition scenarios. It can be seen that $\rho_{4*, \rm{h}}$, $\rho_{4*, \rm{bd}}$, $\rho_{4*, \rm{vl}}$ and $\rho_{4*, \rm{ts}}$ are larger than 1 for all transition scenarios, and significantly larger for most transition scenarios; $\rho_{23, \rm{h}}$ and $\rho_{23, \rm{ts}}$ are smaller than 1 for all transition scenarios, while $\rho_{23, \rm{bd}}$ and $\rho_{23, \rm{vl}}$ are larger than 1 for most transition scenarios; $\rho_{1*, \rm{h}}$, $\rho_{1*, \rm{bd}}$ and $\rho_{1*, \rm{ts}}$ are significantly smaller than 1 for all transition scenarios, and $\rho_{1*, \rm{vl}}$ are smaller than 1 for most transition scenarios. Accordingly, the conclusions \textit{(1)} to \textit{(3)} drawn from the 9-bus system are still valid for most topology transition scenarios of the 118-bus system. 
}

\begin{table}[h]
    \caption{\hig{Average computation time of solving the BTT model}}\label{tab-6-2-5}
    \centering
    \setlength{\tabcolsep}{1.6pt} 
     
    \small{
    \begin{tabularx}{1\linewidth}{p{1.7cm}<{\centering}p{2cm}<{\centering}p{1.7cm}<{\centering}p{1.8cm}<{\centering}p{1.2cm}<{\centering}}
    \hline\hline   
    First-stage BTT model  & AC-feasibility recovery model & Second-stage BTT model & Simulation-based model & Total \\ \hline
    125.37s  & 18.80s & 33.28s &  21.51s& 198.96s \\  \hline\hline
    \multicolumn{5}{@{}p{1\columnwidth}@{}}{
            \footnotesize{\textit{Note}: For each of the last three models, it can be solved in parallel for the $n_{\rm s}$ solutions received. Therefore, the longest computation time when solving the $n_{\rm s}$ solutions is regarded as the computation time of solving the model.}
          }
    \end{tabularx}
    }
\end{table}

\hig{
Additionally, Table \ref*{tab-6-2-5} lists the average computation time of solving the BTT model over the 200 transition scenarios. The average total computation time is 198.96s. This backs the computational suitability of the proposed BTT methodology since this computation time is much shorter than the general execution cycle of OTS. Moreover, the computation time for the first-stage BTT model is the major part. This indicates that in practical applications where a high-quality but not necessarily optimal solution can be sufficient, the total solution time can be significantly reduced by increasing the optimality gap when solving the MISOCP of the first-stage BTT model.
}

\section{Conclusion}

This paper develops a novel and powerful methodology to achieve bumpless topology transition in transmission networks. By optimally adjusting the auxiliary control resources and switching transmission lines, the transition process shows superior bumpiness performance in both static and dynamic aspects. The future work will focus on bumpless topology transition which is finished within a transient time scale, i.e., a few seconds. With high-performance wide area measurement systems, line switching and ACV adjustment can be performed during the transient process, which relax assumption A.2. Thus much more fast and more bumpless topology transition can be promisingly achieved.

\ifCLASSOPTIONcaptionsoff
  \newpage
\fi

\bibliographystyle{IEEEtran}
\bibliography{References/main}

\begin{thebibliography}{10}
\providecommand{\url}[1]{#1}
\csname url@samestyle\endcsname
\providecommand{\newblock}{\relax}
\providecommand{\bibinfo}[2]{#2}
\providecommand{\BIBentrySTDinterwordspacing}{\spaceskip=0pt\relax}
\providecommand{\BIBentryALTinterwordstretchfactor}{4}
\providecommand{\BIBentryALTinterwordspacing}{\spaceskip=\fontdimen2\font plus
\BIBentryALTinterwordstretchfactor\fontdimen3\font minus
  \fontdimen4\font\relax}
\providecommand{\BIBforeignlanguage}[2]{{%
\expandafter\ifx\csname l@#1\endcsname\relax
\typeout{** WARNING: IEEEtran.bst: No hyphenation pattern has been}%
\typeout{** loaded for the language `#1'. Using the pattern for}%
\typeout{** the default language instead.}%
\else
\language=\csname l@#1\endcsname
\fi
#2}}
\providecommand{\BIBdecl}{\relax}
\BIBdecl

\bibitem{4-970}
J.~Li, F.~Liu, Z.~Li, C.~Shao, and X.~Liu, ``Grid-side flexibility of power
  systems in integrating large-scale renewable generations: A critical review
  on concepts, formulations and solution approaches,'' \emph{Renew. Sust.
  Energ. Rev.}, vol.~93, pp. 272--284, Oct. 2018.

\bibitem{4-62}
E.~B. Fisher, R.~P. Oneill, and M.~C. Ferris, ``Optimal transmission
  switching,'' \emph{IEEE Trans. Power Syst.}, vol.~23, no.~3, pp. 1346--1355,
  Aug. 2008.

\bibitem{4-361}
C.~{Li}, H.-D. {Chiang}, and Z.~{Du}, ``Online line switching method for
  enhancing the small-signal stability margin of power systems,'' \emph{IEEE
  Trans. Smart Grid}, vol.~9, no.~5, pp. 4426--4435, Sept. 2018.

\bibitem{4-1369}
E.~Little, S.~Bortolotti, J.-Y. Bourmaud, E.~Karangelos, and Y.~Perez,
  ``Optimal transmission topology for facilitating the growth of renewable
  power generation,'' in \emph{2021 {IEEE} Madrid {PowerTech}}.\hskip 1em plus
  0.5em minus 0.4em\relax {IEEE}, Jun. 2021.

\bibitem{4-859}
T.~Han and D.~J. Hill, ``H2-norm transmission switching to improve synchronism
  of low-inertia power grids,'' \emph{{IFAC}-{PapersOnLine}}, vol.~53, no.~2,
  pp. 13\,299--13\,304, Jul. 2020.

\bibitem{4-1309}
T.~Han, D.~J. Hill, and Y.~Song, ``Optimal topology transition,''
  \emph{arXiv:xxxx.xxxxx}, May 2022.

\bibitem{4-1285}
N.~Martins, E.~J. de~Oliveira, W.~C. Moreira, J.~L.~R. Pereira, and R.~M.
  Fontoura, ``Redispatch to reduce rotor shaft impacts upon transmission loop
  closure,'' \emph{IEEE Trans. Power Syst.}, vol.~23, no.~2, pp. 592--600, May
  2008.

\bibitem{4-558}
G.~M. Huang, W.~Wang, and J.~An, ``Stability issues of smart grid transmission
  line switching,'' \emph{IFAC Proceedings Volumes}, vol.~47, no.~3, pp.
  7305--7310, 2014.

\bibitem{4-161}
R.~Owusu-Mireku and H.-D. Chiang, ``On the dynamics and transient stability of
  power systems post-transmission switching,'' in \emph{Power \& Energy Society
  General Meeting}.\hskip 1em plus 0.5em minus 0.4em\relax IEEE, 2017, pp.
  1--5.

\bibitem{4-1286}
W.~T. Elsayed, H.~Farag, H.~H. Zeineldin, and E.~F. El-Saadany, ``Dynamic
  transitional droops for seamless line-switching in islanded microgrids,''
  \emph{IEEE Trans. Power Syst.}, vol.~36, no.~6, pp. 5590--5601, Nov. 2021.

\bibitem{4-1287}
E.~Pompodakis, G.~C. Kryonidis, and M.~Alexiadis, ``A three-phase
  sensitivity-based approach for smooth line-switching in islanded
  microgrids,'' \emph{techrxiv:14852568}, Jun. 2021.

\bibitem{4-1294}
T.~Han and D.~J. Hill, ``Dispatch of virtual inertia and damping: Numerical
  method with {SDP} and {ADMM},'' \emph{Int. J. Electr. Power Energy Syst.},
  vol. 133, p. 107259, Dec. 2021.

\bibitem{4-1303}
R.~K.~V. R.~Mohan~Mathur, \emph{Thyristor-Based FACTS Controllers for
  Electrical Transmission Systems}.\hskip 1em plus 0.5em minus 0.4em\relax
  Wiley-IEEE Press, Feb. 2002.

\bibitem{4-1293}
M.~A. Gonzalez-Salazar, T.~Kirsten, and L.~Prchlik, ``Review of the operational
  flexibility and emissions of gas- and coal-fired power plants in a future
  with growing renewables,'' \emph{Renew. Sustain. Energy Rev.}, vol.~82, pp.
  1497--1513, Feb. 2018.

\bibitem{4-1292}
D.~Hurley, P.~Peterson, and M.~Whited, ``Demand response as a power system
  resource,'' \emph{Synapse Energy Economics Inc}, 2013.

\bibitem{4-1291}
M.~Base, ``On-load tap-changers for power transformers,'' \emph{Regensburg,
  Germany}, 2013.

\bibitem{4-63}
K.~W. Hedman, R.~P. Oneill, E.~B. Fisher, and S.~S. Oren, ``Optimal
  transmission switching with contingency analysis,'' \emph{IEEE Trans. Power
  Syst.}, vol.~24, no.~3, pp. 1577--1586, May 2009.

\bibitem{4-1289}
J.~V. Milanovic, K.~Yamashita, S.~M. Villanueva, S.~Z. Djokic, and L.~M.
  Korunovic, ``International industry practice on power system load modeling,''
  \emph{IEEE Trans. Power Syst.}, vol.~28, no.~3, pp. 3038--3046, Aug. 2013.

\bibitem{4-1108}
A.~Arif, Z.~Wang, J.~Wang, B.~Mather, H.~Bashualdo, and D.~Zhao, ``Load
  modeling{\textemdash}{A} review,'' \emph{IEEE Trans. Smart Grid}, vol.~9,
  no.~6, pp. 5986--5999, Nov. 2018.

\bibitem{4-1301}
K.~Baker, ``Solutions of {DC} {OPF} are never {AC} feasible,'' in
  \emph{Proceedings of the 12th {ACM} International Conference on Future Energy
  Systems}.\hskip 1em plus 0.5em minus 0.4em\relax {ACM}, Jun. 2021.

\bibitem{4-995-ea}
T.~Han, Y.~Song, and D.~J. Hill, ``Ensuring network connectedness in optimal
  transmission switching problems,'' \emph{{IEEE} Trans. Circuits Syst. {II}},
  vol.~68, no.~7, pp. 2603--2607, Jul. 2021.

\bibitem{4-971-3}
\BIBentryALTinterwordspacing
T.~Han, ``Structure-oriented optimization and control,'' 2021. [Online].
  Available: \url{https://github.com/thanever/SOC/tree/master/Btt/data}
\BIBentrySTDinterwordspacing

\bibitem{4-1382}
P.~Demetriou, M.~Asprou, J.~Quiros-Tortos, and E.~Kyriakides, ``Dynamic {IEEE}
  test systems for transient analysis,'' \emph{{IEEE} Systems Journal},
  vol.~11, no.~4, pp. 2108--2117, Dec. 2017.

\bibitem{4-1288}
J.~Yang, N.~Zhang, C.~Kang, and Q.~Xia, ``A state-independent linear power flow
  model with accurate estimation of voltage magnitude,'' \emph{IEEE Trans.
  Power Syst.}, vol.~32, no.~5, pp. 3607--3617, Sep. 2017.

\bibitem{4-1163}
C.~Coffrin and P.~V. Hentenryck, ``A linear-programming approximation of {AC}
  power flows,'' \emph{{INFORMS} Journal on Computing}, vol.~26, no.~4, pp.
  718--734, Nov. 2014.

\bibitem{4-1300}
W.~Wei, ``Tutorials on advanced optimization methods,''
  \emph{arXiv:2007.13545}, Jul. 2020.

\bibitem{4-1297}
J.~P. Vielma and G.~L. Nemhauser, ``Modeling disjunctive constraints with a
  logarithmic number of binary variables and constraints,'' \emph{Math.
  Program.}, vol. 128, no. 1-2, pp. 49--72, Jul. 2009.

\end{thebibliography}

\appendices
\section*{Appendix: Linearization of the Power Flow Model}\label{appendix-4-1-1}

\subsubsection{Network}
For the network, the AC power flow model is linearized with the decoupled linearized power flow (DLPF) model \cite{4-1288}. 
  The DLPF model is written as
\begin{equation}\label{eq-6-2-7}
            \begin{bmatrix}
                \bm{p}_{\rm fb} \\
                \bm{q}_{\rm fb} 
            \end{bmatrix}
            \!\!=\!
            - \begin{bmatrix}
                \bm{p}_{\rm tb} \\
                \bm{q}_{\rm tb} 
            \end{bmatrix}
            \!\!=\!
            \bm{p}_{\rm z}
            \!\circ\!
            \begin{bmatrix}
                \bm{z} \\
                \bm{z}
            \end{bmatrix}
            \!,
            \bm{p}_{\rm v}
            \!\!=\!\!
            \begin{bmatrix}
                \tilde{\bm{E}} \bm{g}_{\rm lc}\D  \bm{z}  + \bm{g}_{\rm s} \\
                \!-  \tilde{\bm{E}} \bm{b}_{\rm lc}\D \bm{z} \!-\! \bm{b}_{\rm s}
            \end{bmatrix}
            \!\!\circ\!\!
            \begin{bmatrix}
                \bm{v} \\
                \bm{v}
            \end{bmatrix} 
\end{equation}
with
\begin{equation}\label{eq-6-2-7-1}
    \!\!
    \bm{p}_{\rm z} \!\!=\!\! 
    -\!\!
    \begin{bmatrix}
        \bm{b}_{\rm b}\D  \!\!&\!\!\! - \bm{g}_{\rm b}\D \\
        \bm{g}_{\rm b}\D  \!\!&\!\!\! \bm{b}_{\rm b}\D  
    \end{bmatrix}\!\!
    \begin{bmatrix}
       \! \bm{E}\T \bm{\theta} \\
       \! \bm{E}\T \bm{v}
    \end{bmatrix}\!
    \!,
    \bm{p}_{\rm v} \!\!=\!\!\!
    \begin{bmatrix}
        \bm{E}_{\rm g} \bm{p}_{g}  \!-\! \bm{E}_{\rm d} \bm{p}_{\rm d} \\
        \!\bm{E}_{\rm g} \bm{q}_{g}  \!\!-\!\! \bm{E}_{\rm d} \bm{q}_{\rm d} \!\!+\!\! \bm{E}_{\rm c} \bm{q}_{\rm c} \!
    \end{bmatrix}
    \!\!-\!\!
    \begin{bmatrix}
        \! \bm{E} \bm{p}_{\rm fb} \!\! \\
        \! \bm{E} \bm{q}_{\rm fb} \!\!   
    \end{bmatrix} \!\!
\end{equation}

By the Big-M method, (\ref{eq-6-2-7}) is equivalent to 
\begin{adjustwidth}{-0.5em}{}
    \begin{subequations}\label{eq-6-2-7-a}
        \begin{align}
            &
            \!\!-\!\! 
            M\!\!
            \begin{bmatrix}
                \!\bm{1} \!\!-\!\! \bm{z}\! \\
                \!\bm{1} \!\!-\!\! \bm{z}\!
            \end{bmatrix}
            \!\!\leq\!\! 
            \begin{bmatrix}
                \!\bm{p}_{\rm fb}\!\! \\
                \!\bm{q}_{\rm fb}\!\! 
            \end{bmatrix}
            \!-\! \bm{p}_{\rm z}
            \!\!\leq\!  
            M\!
            \begin{bmatrix}
                \!\bm{1} \!\!-\!\! \bm{z}\! \\
                \!\bm{1} \!\!-\!\! \bm{z}\!
            \end{bmatrix}
            \!\!, 
            -\! M\!
            \begin{bmatrix}
                \! \bm{z}\! \\
                \! \bm{z}\!
            \end{bmatrix}
            \!\!\!\leq\!\!\! 
            \begin{bmatrix}
                \!\bm{p}_{\rm fb} \!\! \\
                \!\bm{q}_{\rm fb} \!\! 
            \end{bmatrix}
            \!\!\leq\!
            M\!\!
            \begin{bmatrix}
                \! \bm{z}\! \\
                \! \bm{z}\!
            \end{bmatrix}
            \\&
            [\bm{v} \bm{1}_{n_{\rm e}}\T \!-\! M(\bm{J} \!-\! \bm{1}_{n_{\rm n}} \! \bm{z}\T)] \!\circ\! \tilde{\bm{E}} \leq \bm{U} \leq  M \bm{1}_{n_{\rm n}} \bm{z}\T \!\circ\! \tilde{\bm{E}}
            \\& 
            - M \bm{1}_{n_{\rm n}} \bm{z}\T \!\circ\! \tilde{\bm{E}} \leq \bm{U} \!\leq\!  [\bm{v} \bm{1}_{n_{\rm e}}\T \!+\! M(\bm{J} \!-\! \bm{1}_{n_{\rm n}} \! \bm{z}\T)] \!\circ\! \tilde{\bm{E}} 
            \\&
            \bm{p}_{\rm v}
            =
            \begin{bmatrix}
                \bm{U} \bm{g}_{\rm lc}\D \bm{1}_{n_{\rm e}}   \\
                -  \bm{U} \bm{b}_{\rm lc}\D  \bm{1}_{n_{\rm e}}  
            \end{bmatrix} 
            \!\!+\!\!
            \begin{bmatrix}
                \! \bm{g}_{\rm s}   \!\! \\
                \! - \bm{b}_{\rm s} \!\!  
            \end{bmatrix}
            \!\!\circ\!\!
            \begin{bmatrix}
               \! \bm{v} \!\! \\
               \! \bm{v} \!\!
            \end{bmatrix}\!, 
            \begin{bmatrix}
                \bm{p}_{\rm fb} \\
                \bm{q}_{\rm fb} 
            \end{bmatrix}
            \!=
            \!-\! \begin{bmatrix}
                \bm{p}_{\rm tb} \\
                \bm{q}_{\rm tb} 
            \end{bmatrix}
        \end{align}
    \end{subequations}
\end{adjustwidth}
where $\bm{U} \!\in\! \mathbb{R}^{n_{\rm n} \times n_{\rm e}}$ is an auxiliary variable matrix. 

Since admittances of the branches equipped with TCSC are variables, $\bm{p}_{\rm z}$ still contains bilinear terms, which are further linearized. For any line $e \!\in\! \mathcal{E}$ installed a TCSC, the bilinear terms are $b_{{\rm b}, e} \theta_{ij(e)}$, $g_{{\rm b}, e} \theta_{ij(e)}$, $b_{{\rm b}, e} v_{ij(e)}$, and $g_{{\rm b}, e} v_{ij(e)}$, with $\theta_{ij(e)} \!\!=\! \theta_{i(e)} \!\!-\!\! \theta_{j(e)}$, $v_{ij(e)} \!\!=\! v_{i(e)} \!\!-\!\! v_{j(e)}$, and $i(e)$ and $j(e)$ being the from bus and to bus of branch $e$, respectively. 
A vital point to linearize these terms is to capture the coupling between $b_{{\rm b}, e}$ and $g_{{\rm b}, e}$ caused by the control variable of TCSC, i.e., the reactance of branch $e$. This indicates that discretization based linearization techniques are more suitable here since they can handily tackle this coupling. 
Taking $b_{{\rm b}, e} \theta_{ij(e)}$ and $g_{{\rm b}, e} \theta_{ij(e)}$ for example, $b_{{\rm b}, e} \theta_{ij(e)}$ is first linearized as
\begin{adjustwidth}{-0.5em}{}
\begin{subequations}
    \begin{align} 
    & b_{{\rm b}, e} \theta_{ij(e)}  \!\!=\! 
    b_{{\rm b}, e}\B  \theta_{ij(e)}
    + \bm{\mu}\T \bm{\tau}_{\rm t},  
    \bm{\eta}\T \bm{\tau}_{\rm t} + b_{{\rm b}, e}\B \leq b_{{\rm b}, e}\U
    \\
    & \Delta \theta_e\B \bm{\eta} \!\!\leq\!\! \bm{\mu} \!\!\leq\!\! \Delta \theta_e\U \bm{\eta}
    , \Delta \theta_e\B (\bm{1} \!\!-\!\! \bm{\eta}) \!\!\leq\!\!  \theta_{ij(e)} \bm{1} \!-\! \bm{\mu} \!\leq\! \Delta \theta_e\U (\bm{1} \!-\! \bm{\eta})
    \\
    & \bm{\eta} \in \mathbb{B}^{n_{\rm t}+1}, \bm{\mu} \in \mathbb{R}^{n_{\rm t}+1}
    \end{align}
\end{subequations}
\end{adjustwidth}
where $\bm{\eta} \!\!=\!\! [\eta_k]_{k=0}^{n_{\rm t}} $, 
$\bm{\tau}_{\rm t} \!\!=\!\! [b_{{\rm b}, e}^0 \!\!-\!\! b_{{\rm b}, e}\B~  
2^{k\!-\!1} \frac{b_{{\rm b}, e}\U \!- b_{{\rm b}, e}\B }{2^{n_{\rm t}}}  
]_{k\!=\!1}^{n_{\rm t}} $, $\Delta \theta_e\B$ and $\Delta \theta_e\U$ are the lower and upper bounds of the angle difference of branch $e$, respectively. The accuracy of the linearization can be improved by increasing the value of $n_{\rm t}$, which yields an exponential growth of the number of breakpoints. This number is between $2^{n_{\rm t}}$ and $2^{n_{\rm t} + 1}$, depending on the value of $b_{{\rm b}, e}^0$. 
In addition, when $b_{{\rm b}, e}^0 \!-\! b_{{\rm b}, e}\B$ is exactly an integer multiple of ${(b_{{\rm b}, e}\U \!- b_{{\rm b}, e}\B)}/{2^{n_{\rm t}}}$ except for $b_{{\rm b}, e}\U \!- b_{{\rm b}, e}\B$, the first elements of $\bm{\eta}$ and $\bm{\mu}$ can be removed since they are redundant.

To linearize $g_{{\rm b}, e} \theta_{ij(e)}$, denote by $n_{\rm bp}$ the number of breakpoints of $b_{{\rm b}, e}$, $\hat{\bm{g}}$ the vector of values of $g_{{\rm b}, e}$ corresponding to the breakpoints of $b_{{\rm b}, e}$, and $\hat{g}\B$ and $\hat{g}\U$ the minimum and maximal entry value of $\hat{\bm{g}}$, respectively. Let $\bm{E}_{\rm b} \!\in\! \mathbb{R}^{n_{\rm bp} \times (n_{\rm t} + 1)}$ be the matrix whose rows and columns are associated with breakpoints and elements of $\bm{\eta}$, respectively, and the $i$-th row equals to the value of $\bm{\eta}$ corresponding to the $i$-th breakpoint of $b_{{\rm b}, e}$. By further introducing variables $\bm{\varrho} \in \mathbb{R}^{n_{\rm bp}}$, we have 
\begin{subequations}
    \begin{align}
        & g_{{\rm b}, e} \theta_{ij(e)} = \bm{1}\T \bm{\varrho} \\
        & (\Delta \theta_e\B \hat{g}\B \!\!-\!\! \Delta \theta_e\U \hat{g}\U)  [\bm{E}_{\rm b} \bm{1}_{n_{\rm t} \!+\! 1} \!+\! (\bm{J} \!\!-\!\! 2 \bm{E}_{\rm b}) \bm{\eta}] \!\!\leq\!\! \bm{\varrho} \!-\! \theta_{ij(e)} \hat{\bm{g}} \\ 
        & (\Delta \theta_e\U \hat{g}\U \!\!-\!\! \Delta \theta_e\B \hat{g}\B)  [\bm{E}_{\rm b} \bm{1}_{n_{\rm t} \!+\! 1} \!+\! (\bm{J} \!\!-\!\! 2 \bm{E}_{\rm b}) \bm{\eta}] \!\!\geq\!\! \bm{\varrho} \!-\! \theta_{ij(e)} \hat{\bm{g}} 
    \end{align}
\end{subequations}
Analogously, $b_{{\rm b}, e} v_{ij(e)}$ and $g_{{\rm b}, e} v_{ij(e)}$ can be linearized, where no extra binary variables need to be introduced and $\bm{\eta}$ is shared.



\subsubsection{Generation}
Since the number of generators is much less than that of branches, we adopt a more accurate though less cheap linearization, i.e., the linear-programming approximation of AC power flow (LPAC for short) with slight modification \cite{4-1163}. Specifically, (\ref{eq-6-2-3}) is linearized as
\begin{adjustwidth}{-1em}{}
\begin{subequations}\label{eq-6-2-10}
    \begin{align}
        & 
        \begin{bmatrix}
            \!p_{{\rm g}, i}\!\! \\
            \!q_{{\rm g}, i}\!\!
        \end{bmatrix}
        \!\!\!=\!
        \hat{v}_{{\rm m}, i} \hat{v}_j \!\!
        \begin{bmatrix}
            \!g_{{\rm c}, i} & \!\!\!\! b_{{\rm c}, i}\! \\
            \!b_{{\rm c}, i} & \!\!\!\! g_{{\rm c}, i}\!
        \end{bmatrix}\!\!\!
        \begin{bmatrix}
           \! \varphi_i \\
           \! \theta_{\!{\rm m}, i}\!-\!\theta_{\!j} \!\!
        \end{bmatrix}
        \!\!-\!\!
        \begin{bmatrix}
            \! g_{{\rm c}, i}  \!\!\\
            \! b_{{\rm cc}, i} \!\!
        \end{bmatrix} 
        \! \hat{v}_{\!j}  (2v_{\!j} \!-\! \hat{v}_{\!j})
        \!\!+\!\! 
        \begin{bmatrix}
            0 \\
            \! q_{{\rm g}, i}^{\scriptscriptstyle \Delta}\!\!
        \end{bmatrix}
        \!\!\!\!\\
        &~
        q_{{\rm g}, i}^{\scriptscriptstyle \Delta} 
        \!=\!  b_{{\rm c}, i} (\hat{v}_{{\rm m}, i} v_j \!+\! \hat{v}_j v_{{\rm m}, i} \!-\! 2 \hat{v}_j \hat{v}_{{\rm m}, i}), \varphi_i \!\!\in\!\! \Phi(\theta_{{\rm m}, i}, \theta_{\!j}, \vartheta_{i}\U )
    \end{align}
\end{subequations}  
\end{adjustwidth}
where $\varphi_i$ is the cosine approximation variable for generator $i$; $\hat{v}_{{\rm m}, i}$ and $\hat{v}_j$ denote estimated values of ${v}_{{\rm m}, i}$ and $v_j$, respectively; and $\Phi(\cdot)$ is the polyhedral outer approximation of cosine function $\cos(\theta_{{\rm m}, i}\!-\!\theta_j)$, given by
\begin{equation}\label{eq-6-2-9}
    \!\!\!\!
    \begin{aligned}
        & \Phi(\delta_{i}, \theta_{j}, \vartheta_{i}\U) = \\
        &  \left\{\!\! 
        \varphi_i \!\!\in\!\! \mathbb{R} \!\left\vert\!
        \begin{aligned}
            & \bm{1}_{n_{\rm p}} \varphi_i \!\leq\! \cos( \bm{\tau} \theta_{{\rm p}, i} \!-\! \bm{1}_{n_{\rm p}} \vartheta_{i}\U ) - \\
            &  ~\sin(\bm{\tau} \theta_{\!{\rm p}, i} \!\!-\!\! \bm{1}_{\!n_{\rm p}} \!\vartheta_{i}\U) \!\!\circ\!\! [\bm{1}_{\!n_{\rm p}} \!(\!\delta_{i} \!-\!\theta_{\!j} \!+\! \vartheta_{i}\U) \!\!-\!\! \bm{\tau} \theta_{\!{\rm p}, i} ] \\
            & 0 \leq \varphi_i \leq 1
        \end{aligned}
        \right.\!\!\!
        \right\}
    \end{aligned}\!\!
\end{equation}
where $n_{\rm p}$ is the number of segments of the cosine approximation, $\theta_{{\rm p}, i} \!=\! 2 \vartheta_{i}\U/(1+n_{\rm p})$, and $\bm{\tau} \!=\! [1~2~\cdots n_{\rm p}]\T$. Analogously, (\ref{eq-6-2-2-1}) and (\ref{eq-6-2-2}) can be linearized.

\subsubsection{Loads}
For loads, (\ref{eq-6-2-4:1}) can be linearized by substituting $v_j^2$ with its first-order taylor expansion at $v_j \!\!=\!\! \hat{v}_j$, i.e., $\hat{v}_j(2v_j \!-\! \hat{v}_j)$, and (\ref{eq-6-2-4:2}) and (\ref{eq-6-2-6}) can be linearized analogously to (\ref{eq-6-2-3}) and (\ref{eq-6-2-2-1}), respectively. 
For (\ref{eq-6-2-5}), we first substitute the term $v_j^2$ with its first-order taylor expansion, which gives
\begin{equation}\label{eq-6-2-11}
    q_{{\rm c}, i} = 2 \hat{v}_j b_{{\rm svc}, i} v_j - \hat{v}_j^2 b_{{\rm svc}, i}
\end{equation}
Since $b_{{\rm svc}, i}$ and $v_j$ both have clear upper and lower bounds determined respectively by operating limitations of SVC and the network, we further linearize $b_{{\rm svc}, i} v_j$ by converting it into a separable form where piece-wise linear approximation for univariate nonlinear function can be performed \cite{4-1300}. Specifically, (\ref{eq-6-2-11}) is linearized as
\begin{equation}\label{eq-6-2-12}
    \begin{aligned}
        & q_{{\rm c}, i} = 2 \hat{v}_j ( \hat{\ell}_{{\rm a}, i} - \hat{\ell}_{{\rm b}, i}) - \hat{v}_j^2 b_{{\rm svc}, i} 
        \\
        & \ell_{{\rm a}, i} = ( b_{{\rm svc}, i} + v_j )/2, \ell_{{\rm b}, i} = ( b_{{\rm svc}, i} - v_j )/2
        \\
        & \ell_{{\rm a}, i} \!=\! \bm{\lambda}_{\rm a}\T \bm{\tau}_{\rm a}, 
          \hat{\ell}_{{\rm a}, i} \!=\! \bm{\lambda}_{\rm a}\T \bm{\tau}_{\rm a}^2, 
          \bm{\lambda}_{\rm a} \!\geq\! \bm{0}, 
          \bm{1}\T \bm{\lambda}_{\rm a} \!=\! 1, 
          \bm{\lambda}_{\rm a} \!\in\! \mathbb{SOS}_2
          \\
        & \ell_{{\rm b}, i} \!=\! \bm{\lambda}_{\rm b}\T \bm{\tau}_{\rm b}, 
          \hat{\ell}_{{\rm b}, i} \!=\! \bm{\lambda}_{\rm b}\T \bm{\tau}_{\rm b}^2, 
          \bm{\lambda}_{\rm b} \!\geq\! \bm{0}, 
          \bm{1}\T \bm{\lambda}_{\rm b} \!=\! 1, 
          \bm{\lambda}_{\rm b} \!\in\! \mathbb{SOS}_2 
    \end{aligned}
\end{equation}
where ${\ell}_{{\rm a}, i}, {\ell}_{{\rm b}, i}, \hat{\ell}_{{\rm a}, i}, \hat{\ell}_{{\rm b}, i} \!\in\! \mathbb{R}$, and $\bm{\lambda}_{\rm a}, \bm{\lambda}_{\rm b} \!\in\! \mathbb{R}^{n_{\rm k} + 1}$ are auxiliary variables; $\bm{\tau}_{\rm a} \!=\! [\ell_{{\rm a}, i}\B \!+\! k(\ell_{{\rm a}, i}\U \!- \ell_{\!{\rm a}, i}\B) n_{\rm k}^{\!-1} ]_{k=0}^{n_{\rm k}} $ with $\ell_{{\rm a}, i}\B \!=\! (b_{{\rm svc}, i}\B \!+\! v_j\B)/2$ and $\ell_{{\rm a}, i}\U \!=\! (b_{{\rm svc}, i}\U \!+\! v_j\U)/2$ being the lower and upper bounds of $\ell_{{\rm a}, i}$, respectively; $\bm{\tau}_{\rm b} \!=\! [\ell_{{\rm b}, i}\B \!+\! k(\ell_{{\rm b}, i}\U \!-\! \ell_{{\rm b}, i}\B)n_{\rm k}^{-1} ]_{k=0}^{n_{\rm k}} $ with $\ell_{{\rm b}, i}\B \!=\! (b_{{\rm svc}, i}\B \!-\! v_j\U)/2$ and $\ell_{{\rm b}, i}\U \!=\! (b_{{\rm svc}, i}\U \!-\! v_j\B)/2$ being the lower and upper bounds of $\ell_{{\rm b}, i}$, respectively; $\mathbb{SOS}_2$ stands for the special ordered set of type 2, indicating a vector
of variables with at most two adjacent ones being able to take nonzero values. The $\mathbb{SOS}_2$ requirement can be formulated by introducing $\lceil \log_2 n_{\rm k} \rceil$ auxiliary binary variables, and we refer the reader to  \cite{4-1300, 4-1297} for more details.

Finally, all the above linearization yields (\ref{eq-6-2-cflinear}).

\end{document}